\documentclass[a4paper]{JHEP3}

\usepackage{amsmath,feynmp,slashed,subfig,verbatim,relsize,cite}
\usepackage{graphicx,epsfig}
\newcommand{\gkk}{g_{KK}}
\newcommand{\mkk}{M_{KK}}
\newcommand{\mq}{m_q}

\newcommand{\warpf}{k}
\newcommand{\compradius}{R_c}

\newcommand{\gpolp}{\varepsilon^{\mu}(p)}
\newcommand{\gpolq}{\varepsilon^{\nu}(q)}
\newcommand{\gpolr}{\varepsilon_{\gkk}^{\rho}(r)}
\newcommand{\gpolrprime}{\varepsilon_{\gkk}^{\rho'}(r)}
\newcommand{\gpolrout}{\varepsilon_{\gkk}^{\rho *}(r)}
\newcommand{\gamp}{F_{\mu\nu\rho}}
\newcommand{\gampq}[1]{F_{\mu\nu\rho}^{(q:#1)}}
\newcommand{\gampg}[1]{F_{\mu\nu\rho}^{(g:#1)}}
\newcommand{\gampgh}[1]{F_{\mu\nu\rho}^{({\rm gh}:#1)}}
\newcommand{\gampmom}{\gamp (p,q)}
\newcommand{\eps}[1]{\epsilon_{#1}}
\newcommand{\met}[1]{\eta_{#1}}
\newcommand{\metu}[1]{\eta^{#1}}
\newcommand{\gfived}{g_5}
\newcommand{\gkkk}{g^{(111)}}
\newcommand{\gkkksq}{g^{(111)^{\text{\larger 2}}}}
\newcommand{\gqq}{g^{(1q)}}

\newcommand{\sglfeynint}{\int_0^1 \!\! dx}
\newcommand{\dblfeynint}{\sglfeynint \int_0^{1-x} \!\!\!\!\!\! dy}

\newcommand{\lint}{\int \frac{d^4 l}{(2\pi)^4}}
\newcommand{\linttrf}[2]{\lint
  \frac{\hbox{Tr}\left[ #1 \right]}{#2}}
\DeclareMathOperator{\li}{Li}

\newcommand{\fmfframevertex}[1]{\fmfframe(0,0.7)(0.25,0.5){#1}}

\newcommand{\fddiagwidth}{4.2}
\newcommand{\fddiagheight}{2.5}
\newcommand{\fdshortdiagheight}{1}

\newcommand{\fermiondiagwidth}{4.2}
\newcommand{\fermiondiagheight}{2.5}
\newcommand{\fermionspace}{1cm}

\newcommand{\gluondiagwidth}{4.2}
\newcommand{\gluondiagheight}{2.5}
\newcommand{\gluonspace}{0.34cm}

\newcommand{\ghostdiagwidth}{4.2}
\newcommand{\ghostdiagheight}{2.5}
\newcommand{\ghostspace}{1cm}

\title{Gluon-initiated production of a Kaluza-Klein gluon 
  in a Bulk Randall-Sundrum model}

\author{Benjamin C.~Allanach$^a$, Farvah Mahmoudi$^b$, Jordan P.~Skittrall$^a$
  and K.~Sridhar$^{c,d}$\\
$^a$Department of
  Applied Mathematics and Theoretical Physics, \\ Centre
  for Mathematical Sciences, Univeristy of Cambridge, \\ Wilberforce Road, Cambridge, CB3~0WA,
  United Kingdom\\
\rule{0pt}{3ex}$^b$Laboratoire de Physique Corpusculaire de Clermont-Ferrand (LPC),\\
Universit\'e Blaise Pascal, CNRS/IN2P3, 63177 Aubi\`ere Cedex, France\\
\rule{0pt}{3ex}$^c$LAPTH, Univ.\ de Savoie, CNRS,\\
BP-110, F-74941, Annecy-le-Vieux, France\\
\rule{0pt}{3ex}$^d$Department of Theoretical Physics, \\ Tata
  Institute of Fundamental Research, \\ Homi Bhabha Road, Bombay
  400005, India\\\rule{0pt}{3ex}E-mail: \email{B.C.Allanach@damtp.cam.ac.uk,
     mahmoudi@in2p3.fr, J.P.Skittrall@damtp.cam.ac.uk, sridhar@theory.tifr.res.in}
}

\abstract{In the Bulk Randall-Sundrum model, the Kaluza-Klein excitations
of the gauge bosons are the primary signatures. In particular, the 
search for the Kaluza-Klein (KK) excitation of the gluon at hadron colliders
is of great importance in testing this model. At the leading order in QCD, 
the production of this KK-gluon proceeds only via $q \bar q$-initial states.
We study the production of KK-gluons from gluon initial states at 
next-to-leading order in QCD\@. We find that, even after including the
sub-dominant KK-gluon loops at this order, the next-to-leading order (NLO) cross-section is
tiny compared to the leading order cross-section and unlikely to
impact the searches for this resonance at hardon colliders.}


\preprint{DAMTP-2009-67 \\ LAPTH-1354/09}

\begin{document}
\begin{fmffile}{loop_diagrams}

\section{Introduction}

\noindent The Randall-Sundrum (RS) model \cite{RaS1999} is
a 5-dimensional model
with the fifth dimension
, a slice of anti-de Sitter spacetime with
strong curvature, compactified (to a size comparable to the
Planck length) on a ${\bf S}^1/{\mathbb{Z}_2}$ orbifold. Two branes
are located at the orbifold fixed points $\phi=0,\ \pi$, 
the Planck brane and the TeV brane, respectively.
The Standard Model fields are localised on the TeV brane 
while gravitons exist in the full five-dimensional spacetime.
The five-dimensional spacetime metric is of the form
\begin{equation}
ds^2 =
e^{-{\warpf}\compradius\phi}\eta_{\mu\nu}dx^{\mu}dx^{\nu}~+~\compradius^2d\phi^2
\, ;
\end{equation} 
${\warpf}$ is a mass scale related to the curvature; 
${\rm exp}(-{\warpf} \compradius\phi)$ is the warp factor which rescales
masses of fields localised on the TeV-brane. 
The electroweak hierarchy $\frac{M_P}{M_{EW}} \sim 10^{15}$ 
can be generated by an exponent of order 30 and thus the model 
provides a solution to the hierarchy problem.
For this to work, the compactification radius $R_c$ should be 
stabilised against quantum fluctuations. This can be done by introducing a bulk 
scalar field which generates a potential that allows for
the stabilisation \cite{Goldberger:1999uk,Goldberger:1999un}. The RS model predicts a discrete
spectrum of Kaluza-Klein (KK) excitations of the graviton and
these couple to the Standard Model fields with a coupling
that is enhanced by the warp factor to be of the order of electroweak
strength. Several collider implications of these graviton
resonances have been studied in the literature \cite{DaHR2000,Sridhar:2001sf,AlOPW2000,AlOPPSW2002}. 

Since the original RS model is a model of gravity in AdS spacetime
it is possible to relate it using the AdS/CFT correspondence 
\cite{Maldacena:1997re} to a dual theory -- a strongly coupled gauge theory 
in four dimensions \cite{ArkaniHamed:2000ds,Rattazzi:2000hs}. In this description, it turns
out that the fields localised on the TeV brane are
TeV-scale composites of the strongly interacting theory making the
RS model dual to a composite SM\@. 
Such a composite theory is unviable: 
the simplest possibility is to modify the
model so that only the Higgs field is localised on the TeV brane
while the rest of the SM fields are in the bulk \cite{Po1999,Gherghetta:2000qt}.

In constructing such variants of the RS model, it is not an easy task to
avoid the constraints coming from flavour hierarchy, electroweak precision
tests and flavour-changing neutral currents \cite{Djouadi:2006rk,AgDMS2003,Agashe:2002bx,Contino:2004vy,Agashe:2004rs,AgPS2004}. 
In particular, in order to avoid an unacceptably large contribution
to the electroweak $T$ parameter an enhanced symmetry in the bulk
like $\mathop{\rm SU}(2)_L \times \mathop{\rm SU}(2)_R \times\mathop{\rm U}(1)_{(B-L)}$ may be required. The
heavier fermions need to be closer to the TeV brane so as
to acquire a large Yukawa coupling through a larger overlap with the Higgs
wavefunction. 
In other words, the profiles of the heavier fermions need
to be peaked closer to the TeV-brane. Conversely, 
the fermions close to the Planck brane will have small Yukawa couplings.
However, while the large Yukawa of
the top demands proximity to the TeV brane, the left-handed
electroweak doublet, $(t, b)_L$, cannot be close to the TeV brane
because that induces non-universal couplings of the $b_L$ to the
$Z$. Such couplings are strongly constrained by $R_b$, the measured branching
ratio of $Z \rightarrow b \bar b$. So the doublet needs to be  
as far away from the TeV brane as allowed by $R_b$, whereas the $t_R$
needs to be localised close to the TeV brane to account for the
large Yukawa of the top. We stress that this is one model realization:
a different profile results, for example, in models that invoke 
other symmetry groups in the bulk \cite{Agashe:2006at,Carena:2006bn}. 
It has been found that in order to avoid huge effects of flavour-changing
neutral currents (FCNCs) and to be consistent with precision tests of 
the electroweak sector, the masses of the KK modes of the gauge
bosons have to be strongly constrained. The resulting 
bounds on the masses of the
KK gauge bosons are found to be in the region of 2-3 TeV \cite{AgDMS2003,Agashe:2002bx,Contino:2004vy,Agashe:2004rs,AgPS2004,Djouadi:2009nb} 
though this bound
can be relaxed by enforcing additional symmetries. A review of the
literature on this subject can be found in reference~\cite{Contino:2006nn}.

The $t_R$ localised close to the TeV brane has an enhanced coupling to
the first KK excitation of the gluon in the bulk
and as a result we expect that $\gkk \to t
\bar{t}$ (where $\gkk$ represents the first 
KK excitation of
the gluon) will be a significant decay mode from a discovery
perspective. The decay also lends itself to identification via spin
determination, since the enhanced coupling to $t_R$ over $t_L$ means
the top quarks from 
KK gluon decays will be
polarised~\cite{LiRW2007}. There will be additional challenges for
identification of the $t\bar{t}$ pairs because they will be highly
boosted in the lab frame, but this channel remains a promising search channel.

From a hadron collider perspective, we are interested in the production process
$pp\to \gkk$. The subprocess $q\bar{q} \to \gkk$ has already been
investigated in some detail for the LHC~\cite{Agashe:2006hk,LiRW2007}, as well as
for the Tevatron~\cite{GuMS2007}, but because in many
models the light
quarks have a relatively suppressed coupling to the $\gkk$, it is
worth also considering the process $gg \to \gkk$, even though this
process is one-loop at leading order. It is this process that we
consider in this paper. (Other, tree-level, processes, involving $gg$
fusion to $\gkk$ in association with additional top quark production, have
also been considered 
previously~\cite{GuMS2007a}. A preliminary study of $gg \to \gkk$ has already
been performed~\cite{DjMS2007}, from which our analysis differs by the
consideration of an 
additional channel.)

It might appear \emph{prima facie} that an extension of Yang's
theorem~\cite{Ya1950} 
should forbid the on-shell production of a KK~gluon from two 
on-shell Standard Model gluons. However, to apply an extension of the
theorem to deduce that an amplitude involving three spin-one particles
is zero would require the following conditions to be met:
\begin{itemize}
\item The three particles must be on-shell;
\item Two of the particles must be massless;
\item The amplitude must be symmetric under interchange of the
  massless particles.
\end{itemize}
The final condition is not met in the case of our process $gg\to\gkk$, because
the $\mathop{\rm SU}(3)$ structure of the problem means that the amplitude can
contain 
terms that are antisymmetric under interchange of the Standard Model
gluons. More explicitly, an amplitude involving three coloured gauge particles
can always be decomposed into two parts: one proportional to $f_{abc}$ and
the other proportional to $d_{abc}$. Since the $d_{abc}$'s are symmetric
this part of the amplitude goes to zero due to the usual Yang's theorem
arguments. The antisymmetry of the $f_{abc}$ means that this part of
the amplitude picks up an extra sign and survives the
restrictions of the Yang's theorem.
Therefore, Yang's theorem does not forbid the process $gg\to\gkk$.

\subsection{Theoretical strategy}

In Section~\ref{sec:generalform}, an argument regarding the general form that
must be taken by the amplitude, similar to those of
references~\cite{NiP2005,AlSS2007}, shows that it is then sufficient to
consider a small subset of possible diagrams in order to derive the
overall amplitude.
Consideration of the Feynman rules in Section~\ref{sec:feyn} 
will show that our one-loop
process can obtain contributions from a large number of diagrams, with
loops from fermions, 
KK excitations of fermions, and
KK excitations of gluons, as well as in principle requiring
renormalization. 
In Section~\ref{sec:diags}, we show that it is possible to evade
renormalization by 
considering a particular subset of diagrams, suggested by the general form
argument.
KK excitations of gluons above the first will be
mass-suppressed in the loop, and we therefore neglect diagrams containing such
excitations. 

In addition, it is in principle possible to add a Chern-Simons term to
the Lagrangian of the five-dimensional theory. This term
is gauge-dependent, and a particular five-dimensional gauge choice is
required to 
ensure anomaly cancellation in the four-dimensional effective theory
in which we are working. We start 
by considering a gauge that is simpler for the loop calculation, and show {\em
  a posteriori}\/ in Section~\ref{sec:calc} that
the change of gauge required for anomaly cancellation does not alter
any of the diagrams already considered, and produces only
one more diagram (containing a scalar line). This additional diagram
does not alter the square of the on-shell matrix element. We finally estimate
the cross-section for on-shell KK gluon production in Section~\ref{sec:sec},
before concluding in Section~\ref{sec:disc}. Technical information is
relegated to  appendices: in Appendix~\ref{sec:tensors_satisfying_general_form}, we list the
tensors satisfying the general 
form argument, whereas in Appendix~\ref{sec:gggkkfpintegralanalytic}, we evaluate a Feynman parameter
integral analytically. We provide a recipe for how our results may be adapted
to calculate the off-shell KK gluon amplitude in Appendix~\ref{sec:gen}.

\subsection{Notation}

We shall define the incoming gluon momenta to be $p$ and $q$, with
corresponding polarisation tensors $\gpolp$ and $\gpolq$
respectively. The outgoing KK~gluon momentum is $r$, with
corresponding polarisation tensor $\gpolr$. The polarisation tensors
satisfy
\begin{align}
\gpolp p_{\mu} & = 0 \, ,\\
\gpolq q_{\nu} & = 0 \, ,\\
\gpolr r_{\rho} & = 0 \, .
\end{align}
The momenta satisfy the on-shell conditions
\begin{align}
p^2 = q^2 & = 0 \, \label{eq:ggkkonshellpq} ,\\
r^2 & = \mkk^2 \, ,
\end{align}
where $\mkk$ is the mass of the first Kaluza-Klein excitation of the
gluon, given approximately by the solution of~\cite{Po1999}
\begin{equation}
J_0 \left( \frac{\mkk}{k} e^{\warpf \compradius\pi} \right) = 0 \, ,
\end{equation}
with $J_0$ the Bessel function of the first kind of order zero, $\compradius$
the radius of compactification of the extra dimension, and $\warpf$ the
fixed parameter in the warp factor of order the Planck scale.
Four-momentum conservation ($r=p+q$) yields the on-shell $\gkk$ identity
\begin{equation}
2p\cdot q = \mkk^2 \, .
\end{equation}
Finally, we factor out the polarisation vectors 
to define the tensor $\gampmom$ in terms of the matrix element $\mathcal{M}(p,q)$:
\begin{equation}
\mathcal{M}(p,q) = \gpolrout \gpolp \gpolq \gampmom \, .
\end{equation}

\section{General form of the amplitude\label{sec:generalform}}

We may simplify the calculation by deriving a general form that must
be taken by the amplitude we are calculating.

QCD current conservation, implied by gauge invariance, results in the
properties
\begin{align}
p^{\mu}\gamp & = 0 \, , \label{eq:ggpwardid}\\
q^{\nu}\gamp & = 0 \, . \label{eq:ggqwardid}
\end{align}
Since there are no tree-level diagrams, we may expand $\gamp$ about $p=0$ and
about $q=0$. Expanding first about $p=0$, we may write
\begin{equation}
\gamp =
\mathcal{T}^0_{\mu\nu\rho}(q)+p^{\alpha}\mathcal{T}^1_{\mu\nu\rho\alpha}(p,q)
\, .
\end{equation}
Since equation~\eqref{eq:ggpwardid} must be satisfied for all $p$ with
$|p_0|\leq \mkk$ in the centre of mass frame and $p^2=0$, we can
deduce that
\begin{equation}
\mathcal{T}^0_{\mu\nu\rho} = 0 
\end{equation}
and
\begin{equation}
\mathcal{T}^1_{\mu\nu\rho\alpha} = -\mathcal{T}^1_{\alpha\nu\rho\mu} \, .
\end{equation}
We may therefore write the amplitude as
\begin{equation}
\mathcal{M} = \gpolrout \gpolq (\gpolp p^{\alpha} -
\varepsilon^{\alpha}(p)p^{\mu}) \mathcal{T}^1_{\mu\nu\rho\alpha} \, .
\end{equation}
We next expand about $q=0$, writing
\begin{equation}
\mathcal{T}^1_{\mu\nu\rho\alpha} = t^0_{\mu\nu\rho\alpha}(p) +
q^{\beta}t^1_{\mu\nu\rho\alpha\beta}(p,q) \, .
\end{equation}
Similarly to the previous expansion, we note that since
equation~\eqref{eq:ggqwardid} must be satisfied for all $q$ with
$|q_0|\leq \mkk$ in the centre of mass frame and $q^2=0$, we can
deduce that
\begin{equation}
t^0_{\mu\nu\rho\alpha} = 0
\end{equation}
and
\begin{equation}
t^1_{\mu\nu\rho\alpha\beta} = -t^1_{\mu\beta\rho\alpha\nu} \, .
\end{equation}
We may therefore write the amplitude as
\begin{equation}
\mathcal{M} = \gpolrout (\gpolq q^{\beta} -
\varepsilon^{\beta}(q)q^{\nu}) (\gpolp p^{\alpha} -
\varepsilon^{\alpha}(p)p^{\mu}) t^1_{\mu\nu\rho\alpha\beta}(p,q) \, ,
\label{eq:gggeneralformtensor}\end{equation}
where $t^1_{\mu\nu\rho\alpha\beta}$ has the following properties:
\begin{itemize}
\item Does not contain $q_{\beta}$, $q_{\nu} \,$;
\item Does not contain $p_{\alpha}$, $p_{\mu} \,$;
\item Antisymmetric under $\alpha \leftrightarrow \mu\,$;
\item Antisymmetric under $\beta \leftrightarrow \nu\,$;
\item We may exchange $p_{\rho} \leftrightarrow -q_{\rho}$ in any term.
\end{itemize}
There are $27$ different combinations of $p$, $q$, the metric tensor
$\eta$, and the Levi-Civita tensor $\epsilon$ that have these
properties. The combinations are listed in
Appendix~\ref{sec:tensors_satisfying_general_form}. Having constructed
such terms, we can consider the contribution to $\gamp$ that each will
provide. We note that, by construction, contraction according to
equation~\eqref{eq:gggeneralformtensor} is equivalent to contraction
of the contribution to $t^1_{\mu\nu\rho\alpha\beta}$ with
$\gpolr\gpolq q^{\beta}\gpolp p^{\alpha}$, and we may therefore
contract each contribution according to the equation
\begin{equation}
\gamp = p^{\alpha}q^{\beta}t^1_{\mu\nu\rho\alpha\beta} \, .
\end{equation}
Such a contraction yields four different forms that may contribute to
$\gamp$, namely
\begin{gather}
 \left(\met{\mu\nu}p\cdot q - q_{\mu}p_{\nu}\right)p_{\rho} \, ,\label{eq:gggkkgenformone}\\
\eps{\mu\nu\gamma\delta}p^{\gamma}q^{\delta}p_{\rho} \, ,\label{eq:gggkkgenformtwo} \\
\eps{\mu\nu\rho\gamma}p^{\gamma}p\cdot q -
  \eps{\mu\rho\gamma\delta}p^{\gamma}q^{\delta}p_{\nu}\, ,\\
\eps{\mu\nu\rho\gamma}q^{\gamma}p\cdot q -
  \eps{\nu\rho\gamma\delta}p^{\gamma}q^{\delta}q_{\mu} \, .
\end{gather}
We may therefore write
\begin{multline}
\gamp = A\left(\met{\mu\nu}p\cdot q - q_{\mu}p_{\nu}\right)p_{\rho} +
B\eps{\mu\nu\gamma\delta}p^{\gamma}q^{\delta}p_{\rho} + \\
+ C\left(\eps{\mu\nu\rho\gamma}p^{\gamma}p\cdot q -
  \eps{\mu\rho\gamma\delta}p^{\gamma}q^{\delta}p_{\nu}\right) +
  D\left(\eps{\mu\nu\rho\gamma}q^{\gamma}p\cdot q -
  \eps{\nu\rho\gamma\delta}p^{\gamma}q^{\delta}q_{\mu}\right)
  ,\label{eq:gggkkgeneralform}
\end{multline}
where $A$, $B$, $C$ and $D$ are constants. The problem of calculating
the amplitude reduces to the problem of calculating $A$, $B$, $C$ and
$D$, and where two terms have the same coefficient, it suffices to
evaluate the coefficient for one of them. This last observation will
prove important later for simplifying the calculation.

\section{Feynman rules applicable to the calculation \label{sec:feyn}}

Many of the vertices and propagators applicable to the calculation are
ones that appear in the Standard Model; however, it is necessary to
add to the Feynman rules for those vertices and propagators rules for
the KK~gluon propagator and for its interactions.

Following references~\cite{DaHR1999a,Po1999}, we recall that the
KK~gluon arises from a mode expansion of the components in the usual
four dimensions of the five-dimensional gluon field
into fields depending upon the standard four-dimensional coordinates
($A_{\mu}^{(n)}(x^{\mu})$) and
fields depending upon the extra-dimensional coordinate
($\chi_n(\phi)$). That is (following the conventions of
reference~\cite{DaHR1999a}),
\begin{equation}
A_{\mu}(x^{\mu},\phi) = \sum_{n=0}^{\infty}A_{\mu}^{(n)}(x^{\mu})
\frac{\chi_n(\phi)}{\sqrt{\compradius}} \, ,\label{eq:gmodeexpansion}
\end{equation}
where $\phi = x^4/\compradius$.
The particle to
which we refer as the ``KK~gluon'' is the first excited mode
$A_{\mu}^{(1)}$ -- there
are further excited modes, which we neglect as being suppressed by
their higher masses. The mode decomposition leaves open in principle
the possibility of a scalar gluon $A_4$, corresponding to the
extra-dimensional component of the five-dimensional gluon. We shall
eventually be constrained in our choice of gauge for $A_4$ by 
the requirement that the four-dimensional effective theory be
anomaly-free\footnote{The five-dimensional theory is gauge-variant and
consequently UV-divergent~\cite{AnBDK2006}, but we can impose
upon the four-dimensional theory a condition of gauge-invariance (or
equivalently freedom from gauge anomalies) by
our choice of five-dimensional gauge.}. However, we may begin by making
the gauge choice $A_4=0$ for calculational convenience. Whilst the
subsequent change of gauge affects some Feynman rules already used,
the effects are loop-suppressed, so that their consideration is only
necessary at two-loop level, and may be neglected at the one-loop level.
Oscillations between KK modes are
prevented by the orthonormality condition 
\begin{equation}
\int_{-\pi}^{\pi} d\phi \, \chi_m \chi_n = \delta^{mn}
\, \label{eq:gkkorthogonality}.
\end{equation}
It
is, however, notable that since there need be no momentum conservation in the
extra dimension (there is not translational invariance), there is no
\emph{a priori} reason why there should not be interaction vertices
between the Standard Model gluon and the KK~gluon.

We may obtain the
couplings at the interaction vertices (and, in particular, determine
whether the couplings are non-zero) by integrating out the
extra-dimensional wavefunctions $\chi_n$ that appear in the
interaction terms. Reference~\cite{DaHR1999a} derives values for the
wavefunctions of
\begin{align}
\chi_0 &= \frac{1}{\sqrt{2\pi}} \, \label{eq:chizerowavefunction},\\
\chi_1 &=
\frac{e^{\warpf\compradius |\phi|}}{N_1}
\left[J_1\left(\frac{\mkk}{\warpf }e^{\warpf \compradius |\phi|}\right) + \alpha_1
  Y_1\left(\frac{\mkk}{\warpf }e^{\warpf \compradius |\phi|}\right)\right] \, \label{eq:chionewavefunction},
\end{align}
where $N_1$ is a normalisation constant, $\alpha_1$ is a constant and
$J_1$ and $Y_1$ are Bessel functions of order~$1$. (The derivation is
for an Abelian theory but holds in the non-Abelian case.) The couplings
are determined by substituting equation~\eqref{eq:gmodeexpansion} into
the interacting part of the action (\emph{viz.}\ $-(1/4)
F_{\mu\nu}(x^{\mu},\phi)F^{\mu\nu}(x^{\mu},\phi)$) and integrating
out the fifth dimensional component of the action $x_4 = \compradius \phi$.

For the Standard Model three-point coupling, this procedure gives us
\begin{equation}
g = \gfived \int_{-\pi}^{\pi}d\phi \frac{\chi_0^3}{\sqrt{\compradius }}
= \frac{\gfived}{\sqrt{2\pi \compradius }} \, \label{eq:g5dg4drelation},
\end{equation}
which we use to determine the relationship between the five-dimensional
coupling $\gfived$ and the Standard Model coupling $g$.

For the $gg\gkk$ coupling, we note that all relevant terms in the
action will yield the integral
\begin{equation}
 3\gfived \int_{-\pi}^{\pi}d\phi \frac{\chi_0^2
   \chi_1}{\sqrt{\compradius }} = %
 \frac{3\gfived}{\sqrt{2\pi}} \int_{-\pi}^{\pi}d\phi \frac{\chi_0
  \chi_1}{\sqrt{\compradius }} =  0 \, ,
\end{equation}
where we have used equations~\eqref{eq:chizerowavefunction}
and~\eqref{eq:gkkorthogonality} respectively for the two equalities. So, as already known, there is no
$gg\gkk$ vertex. By a similar argument, there is no $ggg\gkk$ vertex.

For the $g\gkk\gkk$ coupling, the relevant terms in the action will
yield the integral
\begin{equation}
 3\gfived \int_{-\pi}^{\pi}d\phi \frac{\chi_0
   \chi_1^2}{\sqrt{\compradius }} = %
 \frac{3\gfived}{\sqrt{2\pi}} \int_{-\pi}^{\pi}d\phi \frac{
  \chi_1^2}{\sqrt{\compradius }} = %
 \frac{3\gfived}{\sqrt{2\pi \compradius }} =  3g \, ,
\end{equation}
where we have used equations~\eqref{eq:chizerowavefunction},
\eqref{eq:gkkorthogonality} and~\eqref{eq:g5dg4drelation} respectively
in the equalities. Noting that the symmetry
factor for the $g\gkk\gkk$ vertex is $2!$ rather than the $3!$ of the
$ggg$ vertex, we see that the two vertices have the same
coupling.

The integral is very similar for the $gg\gkk\gkk$ coupling: we obtain
\begin{equation}
 3!(\gfived)^2\int_{-\pi}^{\pi}d\phi \frac{\chi_0^2
   \chi_1^2}{{\compradius }}
  =
 \frac{3!(\gfived)^2}{{2\pi}} \int_{-\pi}^{\pi}d\phi \frac{
  \chi_1^2}{{\compradius }} = %
 \frac{3!(\gfived)^2}{{2\pi \compradius }} =  3!g^2 \, ,
\end{equation}
again using equations~\eqref{eq:chizerowavefunction},
\eqref{eq:gkkorthogonality} and~\eqref{eq:g5dg4drelation} respectively
in the equalities. Noting that the symmetry
factor for the $gg\gkk\gkk$ vertex is $2!\cdot 2!$ rather than the $4!$ of the
$gggg$ vertex, we see that the two vertices have the same coupling.

The remaining gluon interaction vertices that we shall need are the
$\gkk\gkk\gkk$ and 
$g\gkk\gkk\gkk$ vertices (the four-point $\gkk\gkk\gkk\gkk$ vertex is
not required). To obtain these two vertices requires integrating
$\chi_1^3$, which, given~\eqref{eq:chionewavefunction}, is
non-trivial. The approximations given in reference~\cite{DaHR1999a}
are sufficient to allow a numerical
integration~\cite{special:Mathematica,special:CUBA}, which shows the
couplings to be non-zero and of the same orders of magnitude as their
Standard Model counterparts (with $\warpf\compradius =11$, there is an enhancement of
approximately $2.5$ times). Since again the symmetry factors and Lagrangian
multiplicities balance between the vertices and their Standard Model
counterparts, we may write the couplings as $\gkkk$ for the
$\gkk\gkk\gkk$ coupling, and $g\gkkk$ for the $g\gkk\gkk\gkk$
coupling, where $\gkkk$ depends upon the geometry of the extra
dimension, but is approximately $2.5g$ when $\warpf\compradius =11$.

The KK~gluon propagator has the same structure as the Standard Model
gluon propagator, but with a mass term. (We choose the Feynman gauge
throughout.)

A one-loop calculation in Feynman gauge in principle can contain
ghosts in diagrams.  In order to provide the appropriate cancellations
the ghosts must have a mode expansion in which the extra-dimensional
component of the expansion is equal to that of the gluon component,
i.e.\ $\chi_n$. This means that the values of the couplings of the
ghost modes match the values of the couplings of the corresponding
gluon modes. The propagator for the KK~ghost has the same structure as
the Standard Model ghost propagator, but with a mass term.

Finally, we shall need to consider the couplings between quarks and
the KK~gluon. These couplings vary with the bulk profiles of the quarks and
with their handedness; we write the magnitude of the coupling as
$\gqq$ in each case and include a chiral projector in the Feynman rule.

The Feynman rules are summarized in Figure~\ref{fig:feynman_rules}.
$f^{abc}$ are the $\mathop{\rm SU}(3)$ anti-symmetric structure constants where 
$a,b,c,\ldots$ are adjoint $\mathop{\rm SU}(3)$ indices. $t^{a}$ are $\mathop{\rm SU}(3)$ generators
and $A,B,C$ denote fundamental $\mathop{\rm SU}(3)$ indices.
\setlength{\unitlength}{1cm}
%
%
%
%
%
%
\newcommand{\feynmanrulesdiagcaption}[2]{%
\hspace{1.5em}
\subfloat{%
\begin{tabular}{c}
\fmfframevertex{%
\begin{fmfgraph*}(\fddiagwidth,\fddiagheight)
%
%
#1
\end{fmfgraph*}}\\
%
%
$ #2 $
\end{tabular}
}
\hspace{1.5em}
}
%
\newcommand{\feynmanrulesdiagcaptionthreelinetext}[4]{%
\hspace{1.5em}
\subfloat{%
\begin{tabular}{c}
\fmfframevertex{%
\begin{fmfgraph*}(\fddiagwidth,\fddiagheight)
%
%
#1
\end{fmfgraph*}}\\
%
%
$ #2 $ \\
 #3 \\
 #4 \\
\end{tabular}
}
\hspace{1.5em}
}
%
%
\newcommand{\feynmanrulesdiagcaptionthreeline}[4]{%
\hspace{1.5em}
\subfloat{%
\begin{tabular}{c}
\fmfframevertex{%
\begin{fmfgraph*}(\fddiagwidth,\fddiagheight)
%
%
#1
\end{fmfgraph*}}\\
%
%
$ #2 $ \\
$ #3 $ \\
$ #4 $ \\
\end{tabular}
}
\hspace{1.5em}
}
%
%
\newcommand{\feynmanrulesdiagcaptionshort}[2]{%
\hspace{1.5em}
\subfloat{%
\begin{tabular}{c}
\fmfframevertex{%
\begin{fmfgraph*}(\fddiagwidth,\fdshortdiagheight)
%
%
#1
\end{fmfgraph*}}\\
%
%
$ #2 $
\end{tabular}
}
\hspace{1.5em}
}
%
%
\newcommand{\feynmanrulespropagator}[4]{%
\fmfleft{l}
\fmfright{r}
\fmf{#1,label=$ #4 $,l.side=left}{l,r}
\fmfv{label=$ #2 $,l.dist=0.6mm}{l}
\fmfv{label=$ #3 $,l.dist=0.6mm}{r}
}
%
%
\newcommand{\feynmanrulesthreepoint}[9]{%
\fmfsurroundn{v}{3}
\fmf{#1,label=$ #3 $,l.side=left}{v1,i}
\fmf{#4,label=$ #6 $,l.side=left}{v2,i}
\fmf{#7,label=$ #9 $,l.side=left}{i,v3}
\fmfv{label=$ #2 $,l.dist=0.6mm}{v1}
\fmfv{label=$ #5 $,l.dist=0.6mm}{v2}
\fmfv{label=$ #8 $,l.dist=0.6mm}{v3}
}
\begin{figure}[p]
\begin{center}
%
%
%
\feynmanrulesdiagcaptionshort{%
\feynmanrulespropagator{fermion}{A,,i}{B,,j}{p}}%
    {\frac{i\delta^{AB}(\slashed{p} - \mq)_{ji}}{p^2 - \mq^2}}
\hspace{\stretch{1}}
%
%
\feynmanrulesdiagcaptionshort{%
\feynmanrulespropagator{gluon}{a,,\alpha}{b,,\beta}{p}}%
    {\frac{-i\delta^{ab}\metu{\alpha\beta}}{p^2}}
\\
\vspace{2em}
%
%
\feynmanrulesdiagcaptionshort{%
\feynmanrulespropagator{dbl_curly}{a,,\alpha}{b,,\beta}{p}}%
    {\frac{-i\delta^{ab}\metu{\alpha\beta}}{p^2-\mkk^2}}
\hspace{\stretch{1}}
%
%
%
\feynmanrulesdiagcaptionshort{%
\feynmanrulespropagator{dbl_dots}{a}{b}{p}}%
    {\frac{\delta^{ab}}{p^2-\mkk^2}}
\\
\vspace{2em}
%
%
\feynmanrulesdiagcaption{%
\feynmanrulesthreepoint{gluon}{a,,\alpha}{}{fermion}{B,,i}{}{fermion}{C,,j}{}}{-ig(t^a)_{CB}(\gamma^{\alpha})_{ji}}
\hspace{\stretch{1}}
%
%
\feynmanrulesdiagcaptionthreelinetext{%
\feynmanrulesthreepoint{dbl_curly}{a,,\alpha}{}{fermion}{B,,i}{}{fermion}{C,,j}{}}{-i\gqq
  (t^a)_{CB}(\gamma^{\alpha})_{jk}\left((1\pm
  \gamma^5)/2\right)_{ki}}%
{($\pm$ in chiral projector according}{to handedness of quark)}
\\
\vspace{2em}
%
%
\feynmanrulesdiagcaptionthreeline{%
\feynmanrulesthreepoint{gluon}{a,,\alpha}{\leftarrow
  p}{gluon}{b,,\beta}{\searrow q}{gluon}{c,,\gamma}{\nearrow r}}{-gf^{abc}[
    (p-q)^{\gamma}\metu{\alpha\beta} + }{ \phantom{gf^{abc}} +
    (q-r)^{\alpha}\metu{\beta\gamma} + }{ \phantom{gf^{abc}} +
    (r-p)^{\beta}\metu{\gamma\alpha}
  ]}
\hspace{\stretch{1}}
%
%
%
%
\feynmanrulesdiagcaptionthreeline{%
\feynmanrulesthreepoint{gluon}{a,,\alpha}{\leftarrow
  p}{dbl_curly}{b,,\beta}{\searrow q}{dbl_curly}{c,,\gamma}{\nearrow r}}{-gf^{abc}[
    (p-q)^{\gamma}\metu{\alpha\beta} + }{ \phantom{gf^{abc}} +
    (q-r)^{\alpha}\metu{\beta\gamma} + }{ \phantom{gf^{abc}} +
    (r-p)^{\beta}\metu{\gamma\alpha}
  ]}
\end{center}
\end{figure}
\begin{figure}[!p]
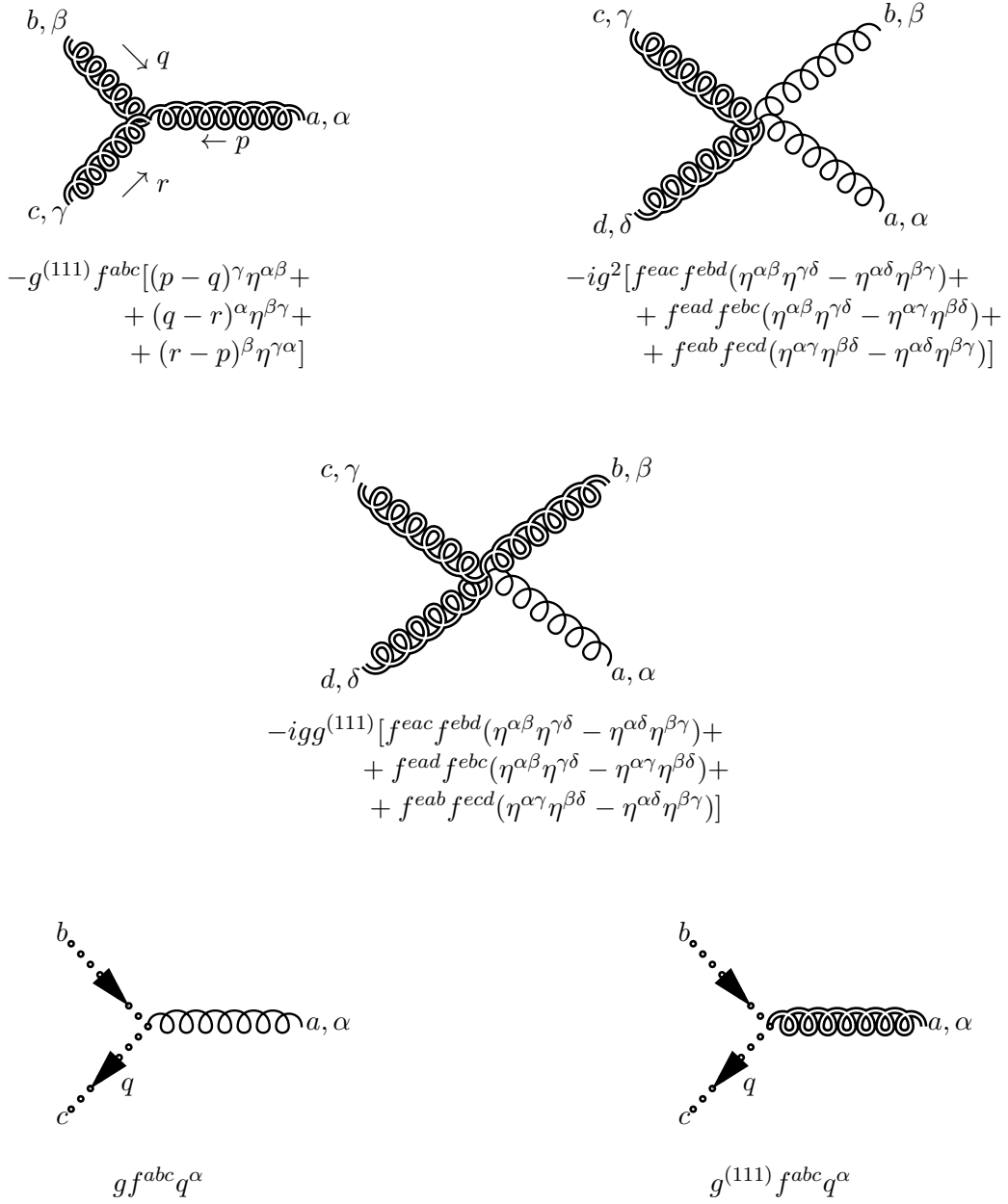

\begin{center}
%
%
%
%
\feynmanrulesdiagcaptionthreeline{%
\feynmanrulesthreepoint{dbl_curly}{a,,\alpha}{\leftarrow
  p}{dbl_curly}{b,,\beta}{\searrow q}{dbl_curly}{c,,\gamma}{\nearrow
  r}}{-\gkkk f^{abc}[
    (p-q)^{\gamma}\metu{\alpha\beta} +}{ \phantom{\gkkk f^{abc}}+
    (q-r)^{\alpha}\metu{\beta\gamma} +}{ \phantom{\gkkk f^{abc}}+
    (r-p)^{\beta}\metu{\gamma\alpha}
  ]}
\hspace{\stretch{1}}
%
%
%
%
%
\feynmanrulesdiagcaptionthreeline{%
\fmfright{v1,v2}
\fmfleft{v4,v3}
\fmf{gluon}{v1,i}
\fmf{gluon}{v2,i}
\fmf{dbl_curly}{v3,i}
\fmf{dbl_curly}{v4,i}
\fmfv{label=$a,,\alpha$,l.dist=0.6mm}{v1}
\fmfv{label=$b,,\beta $,l.dist=0.6mm}{v2}
\fmfv{label=$c,,\gamma$,l.dist=0.6mm}{v3}
\fmfv{label=$d,,\delta$,l.dist=0.6mm}{v4}
}
{-ig^2 [
    f^{eac}f^{ebd}(\metu{\alpha\beta}\metu{\gamma\delta} -
    \metu{\alpha\delta}\metu{\beta\gamma})
  + }{ \phantom{-ig\gkkk \!\!\! } +
    f^{ead}f^{ebc}(\metu{\alpha\beta}\metu{\gamma\delta} -
    \metu{\alpha\gamma}\metu{\beta\delta}) +
}{ \phantom{-ig\gkkk \!\!\! } +
  f^{eab}f^{ecd}(\metu{\alpha\gamma}\metu{\beta\delta} -
  \metu{\alpha\delta}\metu{\beta\gamma})]}
\\
\vspace{2em}
%
%
\feynmanrulesdiagcaptionthreeline{%
\fmfright{v1,v2}
\fmfleft{v4,v3}
\fmf{gluon}{v1,i}
\fmf{dbl_curly}{v2,i}
\fmf{dbl_curly}{v3,i}
\fmf{dbl_curly}{v4,i}
\fmfv{label=$a,,\alpha$,l.dist=0.6mm}{v1}
\fmfv{label=$b,,\beta $,l.dist=0.6mm}{v2}
\fmfv{label=$c,,\gamma$,l.dist=0.6mm}{v3}
\fmfv{label=$d,,\delta$,l.dist=0.6mm}{v4}
}
{-ig\gkkk [
    f^{eac}f^{ebd}(\metu{\alpha\beta}\metu{\gamma\delta} -
    \metu{\alpha\delta}\metu{\beta\gamma})
  + }{ \phantom{-ig\gkkk \!\!\! } +
    f^{ead}f^{ebc}(\metu{\alpha\beta}\metu{\gamma\delta} -
    \metu{\alpha\gamma}\metu{\beta\delta}) +
}{ \phantom{-ig\gkkk \!\!\! } +
  f^{eab}f^{ecd}(\metu{\alpha\gamma}\metu{\beta\delta} -
  \metu{\alpha\delta}\metu{\beta\gamma})]}%
\\
\vspace{2em}
%
%
%
\feynmanrulesdiagcaption{%
\feynmanrulesthreepoint{gluon}{a,,\alpha}{}{dbl_dots_arrow}{b}{}{dbl_dots_arrow}{c}{q}}{gf^{abc}q^{\alpha}}
\hspace{\stretch{1}}
%
%
%
\feynmanrulesdiagcaption{%
\feynmanrulesthreepoint{dbl_curly}{a,,\alpha}{}{dbl_dots_arrow}{b}{}{dbl_dots_arrow}{c}{q}}{\gkkk
  f^{abc}q^{\alpha}}
\caption{Feynman rules required to evaluate the diagrams relevant for
  the $gg\to \gkk$ process. The KK~gluon is denoted by a double gluon
  line, and the KK~ghost is denoted by lines of
  circles.}\label{fig:feynman_rules}
\end{center}
\end{figure}

\clearpage

\section{Diagrams \label{sec:diags}}
\subsection{Quark loop diagrams}
\setlength{\unitlength}{1cm}
%
%
%
%
%
\newcommand{\gggkkfermionedgespace}{1cm}
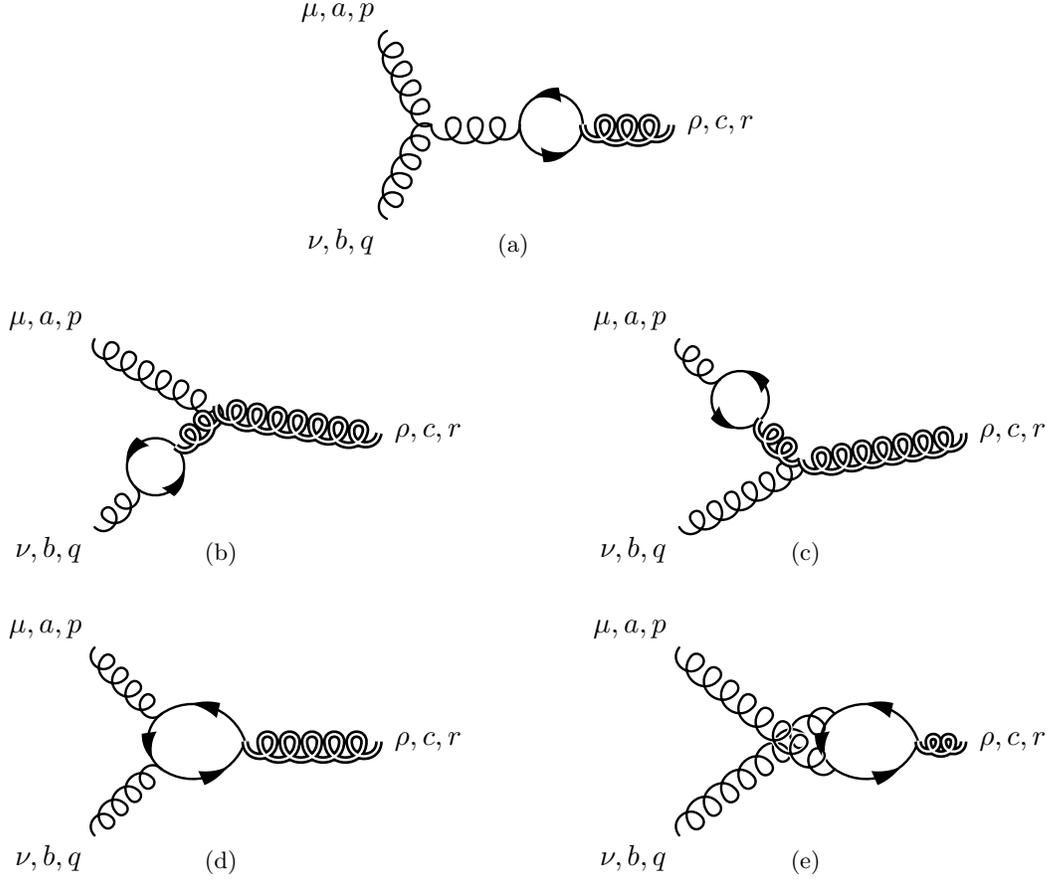
\begin{figure}[!tbhp]
\begin{center}
\subfloat[][]{\begin{fmfgraph*}(\fermiondiagwidth,\fermiondiagheight)
\fmfleft{l1,l2}
\fmfright{r1}
\fmf{gluon}{l2,i1,l1}
\fmf{gluon}{i1,i2}
\fmf{fermion,right,tension=0.7}{i2,i3,i2}
\fmf{dbl_curly}{i3,r1}
\fmflabel{$\mu,a,p$}{l2}
\fmflabel{$\nu,b,q$}{l1}
\fmflabel{$\rho,c,r$}{r1}
\end{fmfgraph*}}\phantom{$\rho,c,r$}\\
\vspace{\fermionspace}
\hspace{\gggkkfermionedgespace}
\subfloat[][]{\begin{fmfgraph*}(\fermiondiagwidth,\fermiondiagheight)
\fmfleft{l1,l2}
\fmfright{r1}
\fmf{gluon}{l1,i1}
\fmf{fermion,right,tension=0.5}{i1,i2,i1}
\fmf{gluon}{l2,i3}
\fmf{dbl_curly}{i2,i3,r1}
\fmflabel{$\mu,a,p$}{l2}
\fmflabel{$\nu,b,q$}{l1}
\fmflabel{$\rho,c,r$}{r1}
\end{fmfgraph*}}
\hspace{\stretch{1}}
\subfloat[][]{\begin{fmfgraph*}(\fermiondiagwidth,\fermiondiagheight)
\fmfleft{l2,l1}
\fmfright{r1}
\fmf{gluon}{l1,i1}
\fmf{fermion,right,tension=0.5}{i1,i2,i1}
\fmf{gluon}{l2,i3}
\fmf{dbl_curly}{i2,i3,r1}
\fmflabel{$\mu,a,p$}{l1}
\fmflabel{$\nu,b,q$}{l2}
\fmflabel{$\rho,c,r$}{r1}
\end{fmfgraph*}}
\hspace{\gggkkfermionedgespace}\phantom{$\rho,c,r$}
\\
\vspace{\fermionspace}
\hspace{\gggkkfermionedgespace}
\subfloat[][]{\begin{fmfgraph*}(\fermiondiagwidth,\fermiondiagheight)
\fmfleft{l1,l2}
\fmfright{r1}
\fmf{gluon}{i1,l1}
\fmf{gluon}{l2,i2}
\fmf{fermion,right=0.5,tension=0.8}{i1,i3,i2,i1}
\fmf{dbl_curly}{i3,r1}
\fmflabel{$\mu,a,p$}{l2}
\fmflabel{$\nu,b,q$}{l1}
\fmflabel{$\rho,c,r$}{r1}
\end{fmfgraph*}}
\hspace{\stretch{1}}
\subfloat[][]{\begin{fmfgraph*}(\fermiondiagwidth,\fermiondiagheight)
\fmfleft{l1,l2}
\fmfright{r1}
\fmf{phantom}{l1,i1}
\fmf{phantom}{l2,i2}
\fmf{fermion,right=0.5,tension=0.8}{i1,i3,i2,i1}
\fmf{dbl_curly}{i3,r1}
\fmffreeze
\fmfshift{(0.28w,0)}{i1,i2,i3}
\fmf{gluon}{i2,l1}
\fmf{gluon,rubout}{l2,i1}
\fmflabel{$\mu,a,p$}{l2}
\fmflabel{$\nu,b,q$}{l1}
\fmflabel{$\rho,c,r$}{r1}
\end{fmfgraph*}}
\hspace{\gggkkfermionedgespace}\phantom{$\rho,c,r$}
\caption{Feynman diagrams for the process that involve a quark in the loop.}
\label{fig:gg-gKK-fermion-loop-diagrams}
\end{center}
\end{figure}

\noindent Figure~\ref{fig:gg-gKK-fermion-loop-diagrams} contains
diagrams for the process that have a quark in the loop. We may write
the contributions to the amplitude from the individual diagrams as
\begin{align}
\gampq{a} & = -\frac{1}{2}i g^2\gqq f^{adb}\delta^{de}
\hbox{Tr}(t^ct^e)\left[ \met{\mu\alpha}(p+r)_{\nu}+ \met{\alpha\nu}
  (-r-q)_{\mu} + \met{\mu\nu}(q-p)_{\alpha}\right] \times \nonumber\\
& \phantom{=-}\times \metu{\alpha\beta}
\frac{1}{r^2}\linttrf{\gamma_{\beta}(\slashed{l}-\mq)\gamma_{\rho}(1\pm\gamma^5)(\slashed{l}+\slashed{r}-\mq)}{[l^2-\mq^2][(l+r)^2-\mq^2]}
\, ,\\
\gampq{b} &= -\frac{1}{2}i g^2\gqq \hbox{Tr}(t^bt^d)\delta^{de}f^{aec}
 \frac{\metu{\alpha\beta}}{q^2-\mkk^2}\left[
   \met{\mu\beta}(p-q)_{\rho} + \met{\beta\rho}(q+r)_{\mu} +
   \met{\rho\mu}(-r-p)_{\beta} \right] \times \nonumber\\
& \phantom{=-}\times
 \linttrf{\gamma_{\nu}(\slashed{l}-\mq)\gamma_{\alpha}(1\pm\gamma^5)(\slashed{l}+\slashed{q}-\mq)}{[l^2-\mq^2][(l+q)^2-\mq^2]}
 \, , \displaybreak[0]\\%
\gampq{c} &= -\frac{1}{2}i g^2\gqq \hbox{Tr}(t^at^d)\delta^{de}f^{bec}
 \frac{\metu{\alpha\beta}}{p^2-\mkk^2}\left[
   \met{\nu\beta}(q-p)_{\rho} + \met{\beta\rho}(p+r)_{\nu} +
   \met{\rho\nu}(-r-q)_{\beta} \right] \times \nonumber\\
& \phantom{=-}\times
 \linttrf{\gamma_{\mu}(\slashed{l}-\mq)\gamma_{\alpha}(1\pm\gamma^5)(\slashed{l}+\slashed{p}-\mq)}{[l^2-\mq^2][(l+p)^2-\mq^2]}
 \, , \displaybreak[0]\\
\gampq{d} & = -\frac{1}{2} g^2 \gqq \hbox{Tr}(t^at^ct^b)\times \nonumber\\
& \phantom{=-}\times
\linttrf{\gamma_{\mu}(\slashed{l}-\slashed{p}-\mq)\gamma_{\rho}(1\pm\gamma^5)(\slashed{l}+\slashed{q}-\mq)\gamma_{\nu}(\slashed{l}-\mq)}{[(l-p)^2-\mq^2][(l+q)^2-\mq^2][l^2-\mq^2]}
\, , \label{eq:gggkkquarkloopdiagd}\\
\gampq{e} & = -\frac{1}{2} g^2 \gqq \hbox{Tr}(t^bt^ct^a)\times \nonumber\\
& \phantom{=-}\times
\linttrf{\gamma_{\nu}(\slashed{l}-\slashed{q}-\mq)\gamma_{\rho}(1\pm\gamma^5)(\slashed{l}+\slashed{p}-\mq)\gamma_{\mu}(\slashed{l}-\mq)}{[(l-q)^2-\mq^2][(l+p)^2-\mq^2][l^2-\mq^2]}
\, ,\label{eq:gggkkquarkloopdiage}
\end{align}
denoting the contribution from quark loop diagram $a$ by $\gampq{a}$,
etc. 
These contributions are applicable to KK~quarks as well, with
appropriate modification of couplings.

\subsection{Kaluza-Klein gluon (and ghost) loop diagrams}

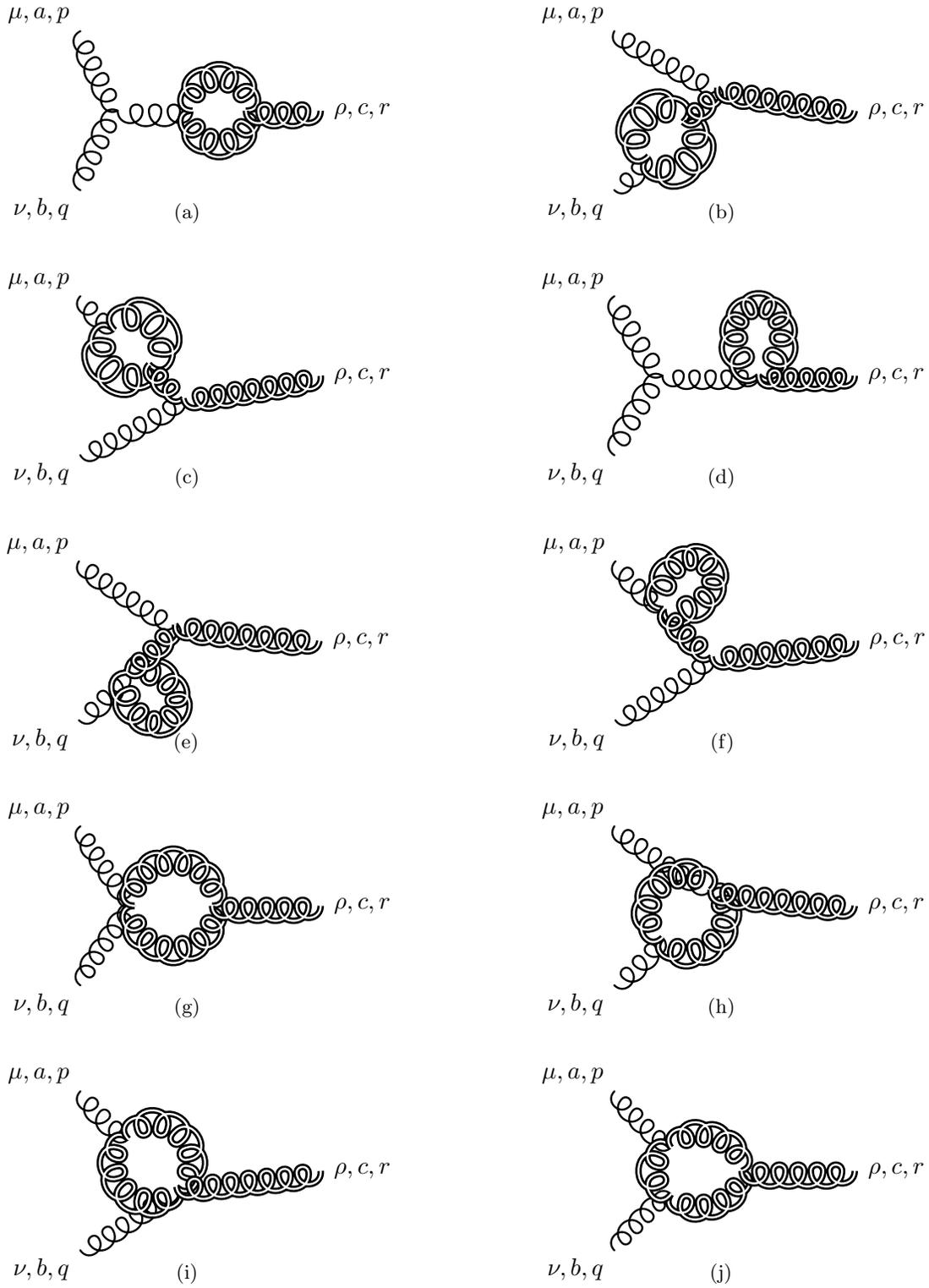
\begin{figure}[!tbhp]
\begin{center}
\phantom{$\nu,b,q$}\hspace{\stretch{0.2}}\subfloat[][]{\begin{fmfgraph*}(\gluondiagwidth,\gluondiagheight)
\fmfleft{l1,l2}
\fmfright{r1}
\fmf{gluon}{l2,i1,l1}
\fmf{gluon}{i1,i2}
\fmf{dbl_curly,right,tension=0.7}{i2,i3,i2}
\fmf{dbl_curly}{i3,r1}
\fmflabel{$\mu,a,p$}{l2}
\fmflabel{$\nu,b,q$}{l1}
\fmflabel{$\rho,c,r$}{r1}
\end{fmfgraph*}}
%
\hspace{\stretch{1}}
\subfloat[][]{\begin{fmfgraph*}(\gluondiagwidth,\gluondiagheight)
\fmfleft{l1,l2}
\fmfright{r1}
\fmf{gluon}{l1,i1}
\fmf{dbl_curly,right,tension=0.5}{i1,i2,i1}
\fmf{gluon}{l2,i3}
\fmf{dbl_curly}{i2,i3,r1}
\fmflabel{$\mu,a,p$}{l2}
\fmflabel{$\nu,b,q$}{l1}
\fmflabel{$\rho,c,r$}{r1}
\end{fmfgraph*}}\hspace{\stretch{0.2}}\phantom{$\rho,c,r$}\\
\vspace{\gluonspace}
%
%
\phantom{$\nu,b,q$}\hspace{\stretch{0.2}}\subfloat[][]{\begin{fmfgraph*}(\gluondiagwidth,\gluondiagheight)
\fmfleft{l2,l1}
\fmfright{r1}
\fmf{gluon}{l1,i1}
\fmf{dbl_curly,right,tension=0.5}{i1,i2,i1}
\fmf{gluon}{l2,i3}
\fmf{dbl_curly}{i2,i3,r1}
\fmflabel{$\mu,a,p$}{l1}
\fmflabel{$\nu,b,q$}{l2}
\fmflabel{$\rho,c,r$}{r1}
\end{fmfgraph*}}
%
\hspace{\stretch{1}}
\subfloat[][]{\begin{fmfgraph*}(\gluondiagwidth,\gluondiagheight)
\fmfleft{l1,l2}
\fmfright{r1}
\fmf{gluon}{l2,i1,l1}
\fmf{gluon}{i1,i2}
\fmf{dbl_curly,right,tension=1}{i2,i2}
\fmf{dbl_curly}{i2,r1}
\fmflabel{$\mu,a,p$}{l2}
\fmflabel{$\nu,b,q$}{l1}
\fmflabel{$\rho,c,r$}{r1}
\end{fmfgraph*}}\hspace{\stretch{0.2}}\phantom{$\rho,c,r$}\\
\vspace{\gluonspace}
%
\phantom{$\nu,b,q$}\hspace{\stretch{0.2}}\subfloat[][]{\begin{fmfgraph*}(\gluondiagwidth,\gluondiagheight)
\fmfleft{l1,l2}
\fmfright{r1}
\fmf{gluon}{l1,i1}
\fmf{dbl_curly,right,tension=0.8}{i1,i1}
\fmf{gluon}{l2,i3}
\fmf{dbl_curly}{i1,i3,r1}
\fmflabel{$\mu,a,p$}{l2}
\fmflabel{$\nu,b,q$}{l1}
\fmflabel{$\rho,c,r$}{r1}
\end{fmfgraph*}}
\hspace{\stretch{1}}
\subfloat[][]{\begin{fmfgraph*}(\gluondiagwidth,\gluondiagheight)
\fmfleft{l2,l1}
\fmfright{r1}
\fmf{gluon}{l1,i1}
\fmf{dbl_curly,right,tension=0.8}{i1,i1}
\fmf{gluon}{l2,i3}
\fmf{dbl_curly}{i1,i3,r1}
\fmflabel{$\mu,a,p$}{l1}
\fmflabel{$\nu,b,q$}{l2}
\fmflabel{$\rho,c,r$}{r1}
\end{fmfgraph*}}\hspace{\stretch{0.2}}\phantom{$\rho,c,r$}\\
\vspace{\gluonspace}
\phantom{$\nu,b,q$}\hspace{\stretch{0.2}}\subfloat[][]{\begin{fmfgraph*}(\gluondiagwidth,\gluondiagheight)
\fmfleft{l1,l2}
\fmfright{r1}
\fmf{gluon}{l2,i1,l1}
\fmf{dbl_curly,right,tension=0.7}{i1,i3,i1}
\fmf{dbl_curly}{i3,r1}
\fmflabel{$\mu,a,p$}{l2}
\fmflabel{$\nu,b,q$}{l1}
\fmflabel{$\rho,c,r$}{r1}
\end{fmfgraph*}}
%
\hspace{\stretch{1}}
\subfloat[][]{\begin{fmfgraph*}(\gluondiagwidth,\gluondiagheight)
\fmfleft{l1,l2}
\fmfright{r1}
\fmf{gluon}{l1,i1}
\fmf{dbl_curly,right,tension=0.5}{i1,i3,i1}
\fmf{gluon}{l2,i3}
\fmf{dbl_curly}{i3,r1}
\fmflabel{$\mu,a,p$}{l2}
\fmflabel{$\nu,b,q$}{l1}
\fmflabel{$\rho,c,r$}{r1}
\end{fmfgraph*}}\hspace{\stretch{0.2}}\phantom{$\rho,c,r$}\\
%
\vspace{\gluonspace}
\phantom{$\nu,b,q$}\hspace{\stretch{0.2}}\subfloat[][]{\begin{fmfgraph*}(\gluondiagwidth,\gluondiagheight)
\fmfleft{l2,l1}
\fmfright{r1}
\fmf{gluon}{l1,i1}
\fmf{dbl_curly,right,tension=0.5}{i1,i3,i1}
\fmf{gluon}{l2,i3}
\fmf{dbl_curly}{i3,r1}
\fmflabel{$\mu,a,p$}{l1}
\fmflabel{$\nu,b,q$}{l2}
\fmflabel{$\rho,c,r$}{r1}
\end{fmfgraph*}}
%
\hspace{\stretch{1}}
\subfloat[][]{\begin{fmfgraph*}(\gluondiagwidth,\gluondiagheight)
\fmfleft{l1,l2}
\fmfright{r1}
\fmf{gluon}{i1,l1}
\fmf{gluon}{l2,i2}
\fmf{dbl_curly,right=0.5,tension=0.8}{i1,i3,i2,i1}
\fmf{dbl_curly}{i3,r1}
\fmflabel{$\mu,a,p$}{l2}
\fmflabel{$\nu,b,q$}{l1}
\fmflabel{$\rho,c,r$}{r1}
\end{fmfgraph*}}\hspace{\stretch{0.2}}\phantom{$\rho,c,r$}\\
\caption{Feynman diagrams for the process that involve a Kaluza-Klein
  gluon in the loop.}
\label{fig:gg-gKK-gluon-loop-diagrams}
\end{center}
\end{figure}

There are no diagrams with a gluon in the loop, since there is no
vertex containing gluons and a single KK~gluon. However, there are
diagrams with KK~gluons (and KK~ghosts) in the loop. These diagrams
are shown in
Figures~\ref{fig:gg-gKK-gluon-loop-diagrams}~and~\ref{fig:gg-gKK-ghost-loop-diagrams}.
\setlength{\unitlength}{1cm}
%
%
%
%

We may write the contributions to the amplitude from the KK~gluon
diagrams as
\begin{align}
\gampg{a} &= -i g^2 \gkkk
f^{abd}f^{efi}f^{chg}\delta^{de}\delta^{fg}\delta^{hi} \left[
  (p-q)_{\alpha}\met{\mu\nu} + (q+r)_{\mu}\met{\nu\alpha} +
  (-r-p)_{\nu}\met{\mu\alpha} \right] \times \nonumber\\
 & \phantom{=} \times \frac{1}{r^2} \metu{\alpha\beta}
\metu{\gamma\delta} \metu{\epsilon\zeta} 
\lint
\Bigg\{ \frac{1}{[(l+r)^2-\mkk^2][l^2-\mkk^2]} \times
\nonumber\\  & \phantom{= \times \frac{1}{r^2} \metu{\alpha\beta}
\metu{\gamma\delta} \metu{\epsilon\zeta} 
\lint} \times[(2r+l)_{\zeta}\met{\beta\gamma} 
  +
    (-2l-r)_{\beta}\met{\gamma\zeta} + (l-r)_{\gamma}\met{\beta\zeta}]
   \times
\nonumber\\  & \phantom{= \times \frac{1}{r^2} \metu{\alpha\beta}
\metu{\gamma\delta} \metu{\epsilon\zeta} 
\lint} \times[(-r+l)_{\delta}\met{\epsilon\rho} +
    (-2l-r)_{\rho}\met{\epsilon\delta} +
    (l+2r)_{\epsilon}\met{\rho\delta}] \Bigg\}\, , \displaybreak[0]\\
\gampg{b} &= -i g^2 \gkkk
f^{cad}f^{efi}f^{bhg}\delta^{de}\delta^{fg}\delta^{hi} \left[
  (-r-p)_{\alpha}\met{\rho\mu} + (p-q)_{\rho}\met{\mu\alpha} +
  (q+r)_{\mu}\met{\rho\alpha} \right] \times \nonumber\\
 & \phantom{=} \times \frac{1}{q^2-\mkk^2} \metu{\alpha\beta}
\metu{\gamma\delta} \metu{\epsilon\zeta} \lint \Bigg\{
\frac{1}{[(l-q)^2-\mkk^2][l^2-\mkk^2]} \times
\nonumber\\ &  \phantom{= \times \frac{1}{q^2-\mkk^2}\metu{\alpha\beta}
\metu{\gamma\delta} \metu{\epsilon\zeta} 
\lint} \times [(-2q+l)_{\zeta}\met{\beta\gamma} +
  (q-2l)_{\beta}\met{\gamma\zeta} + (l+q)_{\gamma}\met{\beta\zeta}]  \times
\nonumber\\  & \phantom{= \times \frac{1}{q^2-\mkk^2} \metu{\alpha\beta}
\metu{\gamma\delta} \metu{\epsilon\zeta} 
\lint} \times [(q+l)_{\delta}\met{\epsilon\nu} +
  (q-2l)_{\nu}\met{\epsilon\delta} + (l-2q)_{\epsilon}\met{\nu\delta}]
\Bigg\} \, , \displaybreak[0]\\
\gampg{c} &= -i g^2 \gkkk
f^{cbd}f^{efi}f^{ahg}\delta^{de}\delta^{fg}\delta^{hi} \left[
  (-r-q)_{\alpha}\met{\rho\nu} + (q-p)_{\rho}\met{\nu\alpha} +
  (p+r)_{\nu}\met{\rho\alpha} \right] \times \nonumber\\
 & \phantom{=} \times \frac{1}{p^2-\mkk^2} \metu{\alpha\beta}
\metu{\gamma\delta} \metu{\epsilon\zeta} \lint \Bigg\{
\frac{1}{[(l-p)^2-\mkk^2][l^2-\mkk^2]} \times
\nonumber\\ &  \phantom{= \times \frac{1}{p^2-\mkk^2}\metu{\alpha\beta}
\metu{\gamma\delta} \metu{\epsilon\zeta} 
\lint} \times [(-2p+l)_{\zeta}\met{\beta\gamma} +
  (p-2l)_{\beta}\met{\gamma\zeta} + (l+p)_{\gamma}\met{\beta\zeta}]  \times
\nonumber\\  & \phantom{= \times \frac{1}{p^2-\mkk^2} \metu{\alpha\beta}
\metu{\gamma\delta} \metu{\epsilon\zeta} 
\lint} \times [(p+l)_{\delta}\met{\epsilon\mu} +
  (p-2l)_{\mu}\met{\epsilon\delta} + (l-2p)_{\epsilon}\met{\mu\delta}]
\Bigg\} \, ,\displaybreak[0]\\
\gampg{d} &= -i g^2 \gkkk \frac{1}{r^2}
\metu{\alpha\beta}\metu{\gamma\delta} f^{abd}
\left[(p-q)_{\alpha}\met{\mu\nu} + (q+r)_{\mu}\met{\nu\alpha} +
  (-r-p)_{\nu}\met{\mu\alpha}\right]  \delta^{de}\delta^{fg} \times \nonumber\\
& \phantom{=} \times \left[
  f^{xeg}f^{xfc}(\met{\beta\gamma}\met{\delta\rho}-\met{\beta\rho}\met{\gamma\delta})
  + f^{xec}f^{xfg}(\met{\beta\gamma}\met{\delta\rho} -
  \met{\beta\delta}\met{\gamma\rho}) +\right. \nonumber\\
& \phantom{= \times \Big[ } \left. +
  f^{xef}f^{xgc}(\met{\beta\delta}\met{\gamma\rho} -
  \met{\beta\rho}\met{\gamma\delta}) \right] 
\lint \frac{1}{l^2-\mkk^2} \, ,\displaybreak[0]\\
\gampg{e} &= -i g^2 \gkkk \frac{1}{q^2-\mkk^2}
\metu{\alpha\beta}\metu{\gamma\delta} f^{cad}
\left[(-r-p)_{\alpha}\met{\rho\mu} + (p-q)_{\rho}\met{\mu\alpha} +
  (q+r)_{\mu}\met{\rho\alpha}\right]  \times \nonumber\\
& \phantom{=} \times \delta^{de}\delta^{fg} \left[
  f^{xeg}f^{xfb}(\met{\beta\gamma}\met{\delta\nu}-\met{\beta\nu}\met{\gamma\delta})
  + f^{xeb}f^{xfg}(\met{\beta\gamma}\met{\delta\nu} -
  \met{\beta\delta}\met{\gamma\nu}) +\right. \nonumber\\
& \phantom{= \times \delta^{de}\delta^{fg} \Big[ } \left. +
  f^{xef}f^{xgb}(\met{\beta\delta}\met{\gamma\nu} -
  \met{\beta\nu}\met{\gamma\delta}) \right] 
\lint \frac{1}{l^2-\mkk^2} \, ,\displaybreak[0]\\
\gampg{f} &= -i g^2 \gkkk \frac{1}{p^2-\mkk^2}
\metu{\alpha\beta}\metu{\gamma\delta} f^{cbd}
\left[(-r-q)_{\alpha}\met{\rho\nu} + (q-p)_{\rho}\met{\nu\alpha} +
  (p+r)_{\nu}\met{\rho\alpha}\right] \times \nonumber\\
& \phantom{=} \times \delta^{de}\delta^{fg} \left[
  f^{xeg}f^{xfa}(\met{\beta\gamma}\met{\delta\mu}-\met{\beta\mu}\met{\gamma\delta})
  + f^{xea}f^{xfg}(\met{\beta\gamma}\met{\delta\mu} -
  \met{\beta\delta}\met{\gamma\mu}) +\right. \nonumber\\
& \phantom{= \times \delta^{de}\delta^{fg} \Big[ } \left. +
  f^{xef}f^{xga}(\met{\beta\delta}\met{\gamma\mu} -
  \met{\beta\mu}\met{\gamma\delta}) \right] 
\lint \frac{1}{l^2-\mkk^2} \, ,\displaybreak[0]\\
\gampg{g} &= -i g^2 \gkkk f^{cfe}\delta^{de}\delta^{fg}
\metu{\alpha\beta}\metu{\gamma\delta} \times \nonumber\\
& \phantom{=} \times \left[
  f^{xad}f^{xbg}(\met{\mu\nu}\met{\alpha\delta}-\met{\mu\delta}\met{\nu\alpha}
  +
  f^{xag}f^{xbd}(\met{\mu\nu}\met{\alpha\delta}-\met{\mu\alpha}\met{\nu\delta})
  + \right. \nonumber\\ & \phantom{= \times \Big[} \left.
  f^{xab}f^{xdg}(\met{\mu\alpha}\met{\nu\delta}-\met{\mu\delta}\met{\nu\alpha})
  \right] \times \nonumber\\ & \phantom{=} \times \lint
  \frac{[(l-r)_{\beta}\met{\rho\gamma}+(-2l-r)_{\rho}\met{\gamma\beta}
      + (l+2r)_{\gamma}\met{\beta\rho}]}{[(l+r)^2-\mkk^2][l^2-\mkk^2]}
  \, ,\displaybreak[0]\\
\gampg{h} &= -i g^2 \gkkk f^{bfe}\delta^{de}\delta^{fg}
\metu{\alpha\beta}\metu{\gamma\delta} \times \nonumber\\
& \phantom{=} \times \left[
  f^{xcd}f^{xag}(\met{\rho\mu}\met{\alpha\delta}-\met{\rho\delta}\met{\mu\alpha}
  +
  f^{xcg}f^{xad}(\met{\rho\mu}\met{\alpha\delta}-\met{\rho\alpha}\met{\mu\delta})
  + \right. \nonumber\\ & \phantom{= \times \Big[} \left.
  f^{xcb}f^{xag}(\met{\rho\alpha}\met{\mu\delta}-\met{\rho\delta}\met{\mu\alpha})
  \right] \times \nonumber\\ & \phantom{=} \times \lint
  \frac{[(l+q)_{\beta}\met{\nu\gamma}+(q-2l)_{\nu}\met{\gamma\beta}
      + (l-2q)_{\gamma}\met{\beta\nu}]}{[(l-q)^2-\mkk^2][l^2-\mkk^2]}
  \, ,\displaybreak[0]\\
\gampg{i} &= -i g^2 \gkkk f^{afe}\delta^{de}\delta^{fg}
\metu{\alpha\beta}\metu{\gamma\delta} \times \nonumber\\
& \phantom{=} \times \left[
  f^{xcd}f^{xbg}(\met{\rho\nu}\met{\alpha\delta}-\met{\rho\delta}\met{\nu\alpha}
  +
  f^{xcg}f^{xbd}(\met{\rho\nu}\met{\alpha\delta}-\met{\rho\alpha}\met{\nu\delta})
  + \right. \nonumber\\ & \phantom{= \times \Big[} \left.
  f^{xcb}f^{xdg}(\met{\rho\alpha}\met{\nu\delta}-\met{\rho\delta}\met{\nu\alpha})
  \right] \times \nonumber\\ & \phantom{=} \times \lint
  \frac{[(l+p)_{\beta}\met{\mu\gamma}+(p-2l)_{\mu}\met{\gamma\beta}
      + (l-2p)_{\gamma}\met{\beta\mu}]}{[(l-p)^2-\mkk^2][l^2-\mkk^2]}
  \, ,\displaybreak[0]\\
\gampg{j} &= -ig^2 \gkkk
f^{adi}f^{cfe}f^{bhg}\delta^{de}\delta^{fg}\delta^{hi}
\metu{\alpha\beta}\metu{\gamma\delta}\metu{\epsilon\zeta} \times \nonumber\\
& \phantom{=} \times \lint
\frac{1}{[(l+p)^2-\mkk^2][(l-q)^2-\mkk^2][l^2-\mkk^2]} \times \nonumber\\
& \phantom{= \times \lint} \times [(2p+l)_{\zeta}\met{\alpha\mu} +
  (-2l-p)_{\mu}\met{\alpha\zeta} + (l-p)_{\alpha}\met{\mu\zeta}]
\times \nonumber\\ 
& \phantom{= \times \lint} \times
       [(-r+l-q)_{\beta}\met{\rho\gamma}+(q-2l-p)_{\rho}\met{\beta\gamma}
         + (l+p+r)_{\gamma}\met{\rho\beta}]\times \nonumber\\ 
& \phantom{= \times \lint} \times [(q+l)_{\delta}\met{\epsilon\nu} +
         (-2l+q)_{\nu} \met{\epsilon\delta} + (l-2q)_{\epsilon}
         \met{\nu\delta} ] \, \label{eq:gggkkgluonloopdiagj},
\end{align}
denoting the contribution from KK~gluon loop diagram $a$ by $\gampg{a}$,
etc\@. We may write the contributions to the amplitude from the KK~ghost
diagrams as
\setlength{\unitlength}{1cm}
%
%
%
%
\newcommand{\gggkkghostedgespace}{1cm}
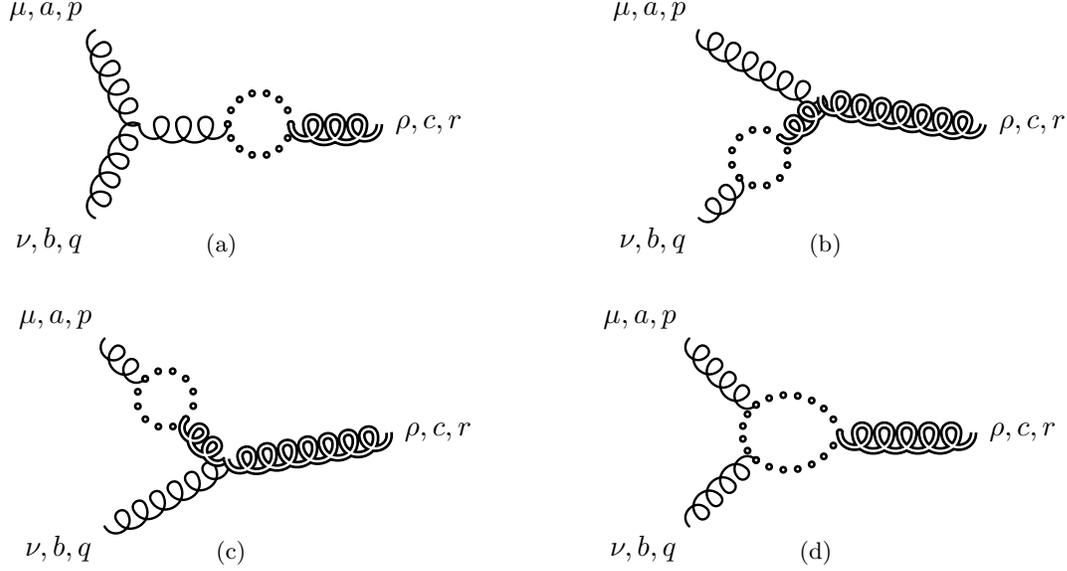
\begin{figure}[tbp]
\begin{center}
\hspace{\gggkkghostedgespace}\subfloat[][]{\begin{fmfgraph*}(\ghostdiagwidth,\ghostdiagheight)
\fmfleft{l1,l2}
\fmfright{r1}
\fmf{gluon}{l2,i1,l1}
\fmf{gluon}{i1,i2}
\fmf{dbl_dots,right,tension=0.7}{i2,i3,i2}
\fmf{dbl_curly}{i3,r1}
\fmflabel{$\mu,a,p$}{l2}
\fmflabel{$\nu,b,q$}{l1}
\fmflabel{$\rho,c,r$}{r1}
\end{fmfgraph*}}
%
%
\hspace{\stretch{1}}
\subfloat[][]{\begin{fmfgraph*}(\ghostdiagwidth,\ghostdiagheight)
\fmfleft{l1,l2}
\fmfright{r1}
\fmf{gluon}{l1,i1}
\fmf{dbl_dots,right,tension=0.5}{i1,i2,i1}
\fmf{gluon}{l2,i3}
\fmf{dbl_curly}{i2,i3,r1}
\fmflabel{$\mu,a,p$}{l2}
\fmflabel{$\nu,b,q$}{l1}
\fmflabel{$\rho,c,r$}{r1}
\end{fmfgraph*}}\hspace{\gggkkghostedgespace}\phantom{$\rho,c,r$}
\\
\vspace{\ghostspace}
\hspace{\gggkkghostedgespace}
%
\subfloat[][]{\begin{fmfgraph*}(\ghostdiagwidth,\ghostdiagheight)
\fmfleft{l2,l1}
\fmfright{r1}
\fmf{gluon}{l1,i1}
\fmf{dbl_dots,right,tension=0.5}{i1,i2,i1}
\fmf{gluon}{l2,i3}
\fmf{dbl_curly}{i2,i3,r1}
\fmflabel{$\mu,a,p$}{l1}
\fmflabel{$\nu,b,q$}{l2}
\fmflabel{$\rho,c,r$}{r1}
\end{fmfgraph*}}
%
\hspace{\stretch{1}}
%
\subfloat[][]{\begin{fmfgraph*}(\ghostdiagwidth,\ghostdiagheight)
\fmfleft{l1,l2}
\fmfright{r1}
\fmf{gluon}{i1,l1}
\fmf{gluon}{l2,i2}
\fmf{dbl_dots,right=0.5,tension=0.8}{i1,i3,i2,i1}
\fmf{dbl_curly}{i3,r1}
\fmflabel{$\mu,a,p$}{l2}
\fmflabel{$\nu,b,q$}{l1}
\fmflabel{$\rho,c,r$}{r1}
\end{fmfgraph*}}
%
%
\hspace{\gggkkghostedgespace}\phantom{$\rho,c,r$}\\
\caption{Feynman diagrams for the process that involve a Kaluza-Klein
  ghost in the loop.}
\label{fig:gg-gKK-ghost-loop-diagrams}
\end{center}
\end{figure}

\begin{align}
\gampgh{a} &= ig^2 \gkkk
f^{adb}f^{cgh}f^{eif}\delta^{de}\delta^{fg}\delta^{hi}
 [(p+r)_{\nu}\met{\mu\alpha} +
  (-r-q)_{\mu}\met{\nu\alpha} + (q-p)_{\alpha}\met{\mu\nu}] \times
\nonumber\\
& \phantom{=} \times \frac{1}{r^2}\metu{\alpha\beta} \lint
\frac{(l+r)_{\rho}l_{\beta}}{[(l+r)^2-\mkk^2][l^2-\mkk^2]} \, ,\displaybreak[0]\\
\gampgh{b} &= ig^2 \gkkk
f^{aec}f^{dif}f^{bgh}\delta^{de}\delta^{fg}\delta^{hi}
[(p-q)_{\rho}\met{\beta\mu} + (q+r)_{\mu}\met{\beta\rho} +
  (-r-p)_{\beta}\met{\rho\mu}] \times
\nonumber\\
& \phantom{=} \times \frac{1}{q^2-\mkk^2}\metu{\alpha\beta} \lint
\frac{l_{\alpha}(l+q)_{\nu}}{[l^2-\mkk^2][(l+q)^2-\mkk^2]} \, ,\displaybreak[0]\\
\gampgh{c} &= ig^2 \gkkk
f^{bec}f^{dif}f^{agh}\delta^{de}\delta^{fg}\delta^{hi}
[(q-p)_{\rho}\met{\beta\nu} + (p+r)_{\nu}\met{\beta\rho} +
  (-r-q)_{\beta}\met{\rho\nu}] \times
\nonumber\\
& \phantom{=} \times \frac{1}{p^2-\mkk^2}\metu{\alpha\beta} \lint
\frac{l_{\alpha}(l+p)_{\nu}}{[l^2-\mkk^2][(l+p)^2-\mkk^2]} \, ,\displaybreak[0]\\
\gampgh{d} &= ig^2 \gkkk
f^{aid}f^{cef}f^{bgh}\delta^{de}\delta^{fg}\delta^{hi} \times
\nonumber\\
& \phantom{=} \times \lint
\frac{(l+p)_{\mu}(l-q)_{\rho}l_{\nu}}{[(l+p)^2-\mkk^2]
  [(l-q)^2-\mkk^2] [l^2-\mkk^2]} \,\label{eq:gggkkghostloopdiagd} ,
\end{align}
denoting the contribution from KK~ghost loop diagram $a$ by $\gampgh{a}$,
etc.

\subsection{Counterterm diagrams}

If we are to use bare parameters in the expressions for the diagrams
given so far, then diagrams containing counterterms will appear to
balance the divergences from the previous diagrams. However, we shall
see that it is possible to avoid considering such diagrams, and so we
shall not detail them in full here. It suffices to observe that there
are four counterterm diagrams (their appearance is that of the
KK~ghost loop diagrams in
Figure~\ref{fig:gg-gKK-ghost-loop-diagrams}, with each KK~ghost loop
replaced by a counterterm), and that the Lorentz structure of the
counterterms is derived from consideration of the underlying
Lagrangian term. In particular, each term in the three-point
counterterm vertex has 
one momentum factor carrying a Lorentz index (the other two being
carried by a metric factor), and the two-point
counterterm vertex has the sum of a term where the metric carries both
external Lorentz indices and a term where there are two momentum terms
each carrying an external Lorentz index. We note that this latter term
always contains a momentum factor that contracts with an external
polarisation vector to give zero. It is therefore the case that none
of the counterterm diagrams contains a term where there is more than
one momentum factor carrying an external Lorentz index. (There is also
no term with a Levi-Civita tensor carrying an external Lorentz index.)

\section{Calculation of the amplitude \label{sec:calc}}

\subsection{Simplification of the calculation}
Before proceeding to calculate the $gg\to\gkk$ amplitude, we note that
we can simplify our calculation significantly by using the general
form derived in equation~\eqref{eq:gggkkgeneralform} to justify
disregarding many diagrams.

Firstly, we note that the only diagrams capable of producing a
Levi-Civita tensor are those containing a trace of a $\gamma^5$,
i.e.\ the diagrams with quark loops. Of the diagrams with quark loops,
we note that the loop integrals for diagrams (a), (b) and (c) only
contain the loop momentum and one other momentum, and have as a maximum two
factors of the momentum on the numerator (both of which contract with
a trace of gamma matrices). This means that, even taking
reparametrisation of the integrand into account, the only possible
terms in the numerator contain either
\begin{itemize}
\item Two identical momenta contracted with a Levi-Civita tensor,
  which gives zero since the Levi-Civita tensor is antisymmetric, or
\item One loop momentum and one other momentum contracted with a
  Levi-Civita tensor, which gives zero since such a term is odd in the
  loop momentum and the loop momentum
  integral is over all of space-time, or
\item No loop momenta, but such a term does not yield a Levi-Civita
  tensor, since the trace involving a $\gamma^5$ term contains only
  two other gamma matrices, and this is zero.
\end{itemize}
So the only contributions to amplitude coefficients $B$, $C$ and $D$
come from $\gampq{d}$ and $\gampq{e}$.

Secondly, we note that in evaluating the contribution to amplitude
coefficient $A$, we may sum the coefficients of either the term
$\met{\mu\nu}p\cdot q p_{\rho}$ or the term
$-q_{\mu}p_{\nu}p_{\rho}$. We choose the latter term.

We have already noted that no counterterm diagram contains more than
one loop momentum carrying an external Lorentz index, so no
counterterm diagram provides a contribution we need to evaluate.

In the quark loop sector, it initially appears that we can obtain a
contribution we need to evaluate from each diagram, it being possible
to obtain terms with three external momenta carrying external Lorentz
indices in each case. However, we note that considering the loop
momentum integral, in the case of diagram~(a) such a term would have
to contain a factor of $r_{\rho}$, which contracts with the external
polarisation vector to give zero, and similarly such a term in
diagram~(b) would have to contain a factor of $q_{\nu}$ and such a
term in diagram~(c) would have to contain a factor of $p_{\mu}$, both
of which contract with external polarisation vectors to give zero. So
the only contributions to amplitude coefficient $A$ from the quark
loop sector that we need to evaluate come from $\gampq{d}$ and $\gampq{e}$.

Similarly, whilst in the KK~gluon loop sector, it initially appears
that we can obtain a contribution we need to evaluate from diagrams
(a), (b), (c) and (j), we note that the terms with three external
momenta carrying external Lorentz indices in diagrams (a), (b) and (c)
contain factors of $r_{\rho}$, $q_{\nu}$ and $p_{\mu}$ respectively,
all of which contract with external polarisation vectors to give
zero. So the only contribution to amplitude coefficient $A$ from the
KK~gluon loop sector that we need to evaluate comes from $\gampg{j}$.

The behaviour of the KK~ghost loop sector is similar to that of the
quark loop sector, and the only contribution to the amplitude
coefficient $A$ from the KK~ghost loop sector that we need to evaluate
comes from $\gampgh{d}$.

We have therefore reduced the calculations required to derive the
amplitude to those required to deduce the coefficients of single
Levi-Civita tensors and of the term $-q_{\mu}p_{\nu}p_{\rho}$ in
$\gampq{d}+\gampq{e}$, and of the term $-q_{\mu}p_{\nu}p_{\rho}$ in
$\gampg{j}+\gampgh{d}$.

\subsection{Contribution from diagrams with quark loops}

We have established that we only need to consider diagrams (d) and (e)
in the quark loop sector, and noting the similarities in their
structure we begin by attempting to sum the diagrams without
evaluating them.

Firstly, we note that we can apply recursively the identity
\begin{equation}
t^at^b = \frac{1}{6} \delta^{ab}I_3 + \frac{1}{2}(if^{abc}+d^{abc})t^c
\, ,
\end{equation}
along with the property that the $t^a$ are traceless, to deduce that
\begin{equation}
\hbox{Tr}(t^at^bt^c) =
\frac{1}{4}(if^{abc}+d^{abc}) \label{eq:sutraceidentity}\, .
\end{equation}
Next, we note that we can apply the charge conjugation relations
\begin{equation}
C^{-1}\gamma^{\mu}C = -\gamma^{\mu \, T} \, , \qquad
C^{-1}\gamma^{\mu}\gamma^5 C = \left( \gamma^{\mu} \gamma^5 \right)^T
\, ,
\end{equation}
along with the cyclic property of the trace and the trace reversal
property of the transpose, to obtain
\begin{multline}
\linttrf{\gamma_{\nu}(\slashed{l}-\slashed{q}-\mq)\gamma_{\rho}(1\pm\gamma^5)(\slashed{l}+\slashed{p}-\mq)\gamma_{\mu}(\slashed{l}-\mq)}{[(l-q)^2-\mq^2][(l+p)^2-\mq^2][l^2-\mq^2]}
=\\
=
\linttrf{(-\slashed{l}-\mq)(-\gamma_{\mu})(-\slashed{l}-\slashed{p}-\mq)\gamma_{\rho}(-1\pm\gamma^5)(-\slashed{l}+\slashed{q}-\mq)(-\gamma_{\nu})}{[(l-q)^2-\mq^2][(l+p)^2-\mq^2][l^2-\mq^2]}
=\\
=
\linttrf{(\slashed{l}-\mq)\gamma_{\mu}(\slashed{l}-\slashed{p}-\mq)\gamma_{\rho}(1\mp\gamma^5)(\slashed{l}+\slashed{q}-\mq)\gamma_{\nu}}{[(l+q)^2-\mq^2][(l-p)^2-\mq^2][l^2-\mq^2]}
\, ,\label{eq:gggkkquarkccidentity}
\end{multline}
where we have taken $l \to -l$ in the final line.

Substituting equations~\eqref{eq:sutraceidentity}
and~\eqref{eq:gggkkquarkccidentity} into
equations~\eqref{eq:gggkkquarkloopdiagd}
and~\eqref{eq:gggkkquarkloopdiage}, we obtain
\begin{align}
\gampq{d}+\gampq{e} & = -\frac{1}{4}g^2 \gqq i f^{acb}
\linttrf{\gamma_{\mu}(\slashed{l}-\slashed{p}-\mq)\gamma_{\rho}(\slashed{l}+\slashed{q}-\mq)\gamma_{\nu}(\slashed{l}-\mq)}{[(l-p)^2-\mq^2][(l+q)^2-\mq^2][l^2-\mq^2]} \mp
\nonumber\\
& \phantom{=} \mp\frac{1}{4}g^2 \gqq d^{acb}
\linttrf{\gamma_{\mu}(\slashed{l}-\slashed{p}-\mq)\gamma_{\rho}\gamma^5(\slashed{l}+\slashed{q}-\mq)\gamma_{\nu}(\slashed{l}-\mq)}{[(l-p)^2-\mq^2][(l+q)^2-\mq^2][l^2-\mq^2]}
\, . \label{eq:gggkkquarkdesum}
\end{align}
Applying a Feynman parametrisation and integral redefinition
\begin{multline}
\lint \frac{f(l)}{[(l-p)^2-\mq^2][(l+q)^2-\mq^2][l^2-\mq^2]} = \\ = 
\dblfeynint \lint \frac{f(l+xp-yq)}{[l^2+2xy \, p\cdot q -
    \mq^2]^3} \, ,
\end{multline}
where we have used equation~\eqref{eq:ggkkonshellpq} ($p^2=q^2=0$) to
simplify the final denominator, and keeping from the first integral of
equation~\eqref{eq:gggkkquarkdesum} only the terms parallel to
$-q_{\mu}p_{\nu}p_{\rho}$ and from the second integral of the equation
only the terms parallel to
$\eps{\mu\nu\gamma\delta}p^{\gamma}q^{\delta}p_{\rho}$,
$\eps{\mu\rho\gamma\delta}p^{\gamma}q^{\delta}p_{\nu}$, and
$\eps{\nu\rho\gamma\delta}p^{\gamma}q^{\delta}q_{\mu}$
(these are the only terms required to evaluate the coefficients $A$,
$B$, $C$ and $D$ in equation~\eqref{eq:gggkkgeneralform}), we
obtain~\cite{special:FORM}
\begin{multline}
\left. \gampq{d}+\gampq{e}
\right|_{\begin{subarray}{l} \text{relevant}
    \\ \text{terms} \end{subarray}} = \\
= 2 i g^2
\gqq f^{abc} \dblfeynint \lint \frac{p_{\nu}q_{\mu} xy
     [p_{\rho}(1-2x) + q_{\rho}(2y-1)]}{[l^2+2xy \, p\cdot q - \mq^2]^3} \pm
\\
\pm 2 i g^2 \gqq d^{abc} \dblfeynint \lint
\frac{xy(\epsilon_{\mu\rho\gamma\delta}p^{\gamma}q^{\delta}p_{\nu} -
  \epsilon_{\nu\rho\gamma\delta}p^{\gamma}q^{\delta}q_{\mu})}{[l^2+2xy
    \, p\cdot
    q - \mq^2]^3} \,
.
\end{multline}
The momentum
integrals may be evaluated by standard techniques to obtain
\begin{multline}
\left. \gampq{d}+\gampq{e}
\right|_{\begin{subarray}{l} \text{relevant}
    \\ \text{terms} \end{subarray}} = \\
= \frac{g^2
\gqq f^{abc}}{(4\pi)^2} \dblfeynint \frac{p_{\nu}q_{\mu} xy
     [p_{\rho}(1-2x) + q_{\rho}(2y-1)]}{\mq^2 - 2xy \, p \cdot q} \pm \\
\pm \frac{g^2 \gqq d^{abc}}{(4\pi)^2}
(\epsilon_{\mu\rho\gamma\delta}p^{\gamma}q^{\delta}p_{\nu} -
\epsilon_{\nu\rho\gamma\delta}p^{\gamma}q^{\delta}q_{\mu}) \dblfeynint
\frac{xy}{\mq^2 - 2xy \, p\cdot q} = \\
= \frac{2 g^2 \gqq f^{abc}}{(4\pi)^2} q_{\mu}p_{\nu}p_{\rho} \, I(\mq, \mkk) \pm  \frac{g^2 \gqq d^{abc}}{(4\pi)^2}
(\epsilon_{\mu\rho\gamma\delta}p^{\gamma}q^{\delta}p_{\nu} -
\epsilon_{\nu\rho\gamma\delta}p^{\gamma}q^{\delta}q_{\mu}) \, K(\mq,
\mkk) \, ,
\end{multline}
where
\begin{align}
I(\mq, \mkk) &= \dblfeynint \frac{xy(1-x-y)}{\mq^2 - xy\mkk^2} \label{eq:gggkkfpnastyintegral}\\
\intertext{and}
K(\mq, \mkk) &=  \dblfeynint \frac{xy}{\mq^2 - xy\mkk^2} \label{eq:gggkkfpnotquiteasnastyintegral}\, .
\end{align}
So each quark loop contributes a total of $-\frac{2 g^2 \gqq
  f^{abc}}{(4\pi)^2}I(\mq, \mkk) $ to the coefficient $A$ in
equation~\eqref{eq:gggkkgeneralform}, a total of $\mp\frac{g^2 \gqq
  d^{abc}}{(4\pi)^2} K(\mq,\mkk)$ to the coefficient $C$ and a total
of  $\pm\frac{g^2 \gqq
  d^{abc}}{(4\pi)^2} K(\mq,\mkk)$ to the coefficient $D$, where the
sign of the contribution varies as the quark is right- or
left-handed. (We shall evaluate the integral $I(\mq,\mkk)$ later; it
will turn out that we shall not need the integral $K(\mq,\mkk)$.)

\subsection{Contribution from diagrams with Kaluza-Klein gluon (or
  ghost) loops}
We have established that we only need to consider diagram~(j) in the
KK~gluon loop sector and diagram~(d) from the KK~ghost loop. It is
useful first to derive an identity for the $\mathop{\rm SU}(3)$ structure constants
contained in the expressions for these diagrams.

The structure constants $f^{abc}$ satisfy the identity
\begin{equation}
(T^a)_{bc} = -if^{abc} \, ,
\end{equation}
where the $T^a$ are in the adjoint representation of $\mathop{\rm SU}(3)$ and
satisfy the same algebra as the fundamental representation. It
therefore follows that
\begin{align}
f^{adh}f^{cfd}f^{bhf} & = -i (T^a)_{dh}(T^c)_{fd}(T^b)_{hf} =
\nonumber\\
 & = -i \hbox{Tr}(T^aT^bT^c) = \nonumber\\
 & = \frac{1}{4}(f^{abc}-id^{abc}) \, ,
\end{align}
using equation~\eqref{eq:sutraceidentity}.

We may now apply a Feynman parametrisation and integral redefinition
to equations~\eqref{eq:gggkkgluonloopdiagj}
and~\eqref{eq:gggkkghostloopdiagd} and evaluate the
numerators~\cite{special:FORM}, obtaining
\begin{align}
\gampg{j} & = - \frac{9}{2} g^2 \gkkk (if^{abc}+d^{abc})
q_{\mu}p_{\nu}p_{\rho} \dblfeynint \lint \frac{xy(1-x-y)}{[l^2+2xy \,
    p\cdot q - \mkk^2]^3} \, ,\\
\gampgh{d} & = \frac{1}{4} g^2 \gkkk (if^{abc}+d^{abc})
q_{\mu}p_{\nu}p_{\rho} \dblfeynint \lint \frac{xy(1-x-y)}{[l^2+2xy \,
    p\cdot q - \mkk^2]^3} \, .
\end{align}
The momentum integrals may be evaluated by standard techniques to
obtain
\begin{align}
\gampg{j} &= -\frac{9}{4} \frac{g^2
  \gkkk}{(4\pi)^2}(f^{abc}-id^{abc})q_{\mu}p_{\nu}p_{\rho} \,
I(\mkk,\mkk) \, ,\\
\gampgh{d} &= \frac{1}{8} \frac{g^2
  \gkkk}{(4\pi)^2}(f^{abc}-id^{abc})q_{\mu}p_{\nu}p_{\rho} \,
I(\mkk,\mkk) \, ,
\end{align}
Where $I(\mkk,\mkk)$ is defined in
equation~\eqref{eq:gggkkfpnastyintegral}.

We therefore obtain a contribution of $\frac{17}{8}\frac{g^2
  \gkkk}{(4\pi)^2}(f^{abc}-id^{abc}) I(\mkk,\mkk)$ to the coefficient $A$ in
equation~\eqref{eq:gggkkgeneralform} from the KK~gluon and KK~ghost
loops.

\subsection{Overall amplitude before anomaly cancellation}
Summing the contributions from the quark loop diagrams and the
KK~gluon and KK~ghost loop diagrams, we derive that the amplitude for
the process $gg\to \gkk$, neglecting higher
order excitations of the KK~gluon and KK~ghost, satisfies
\begin{multline}
\gamp = A\left(\met{\mu\nu}p\cdot q - q_{\mu}p_{\nu}\right)p_{\rho} +
\\
+ C\left(\eps{\mu\nu\rho\gamma}p^{\gamma}p\cdot q - \eps{\mu\nu\rho\gamma}q^{\gamma}p\cdot q -
  \eps{\mu\rho\gamma\delta}p^{\gamma}q^{\delta}p_{\nu} +
  \eps{\nu\rho\gamma\delta}p^{\gamma}q^{\delta}q_{\mu}\right) , \label{eq:gggkkgeneralformsimplified}
\end{multline}
where
\begin{align}
A &= \frac{17}{8}\frac{g^2
  \gkkk}{(4\pi)^2}(f^{abc}-id^{abc}) I(\mkk,\mkk) -\sum_{q_L,q_R} \frac{2 g^2 \gqq
  f^{abc}}{(4\pi)^2}I(\mq, \mkk)\, , \label{eq:gggkkgeneralformexplicit}\\
C &= \sum_{q_R}\frac{g^2 \gqq
  d^{abc}}{(4\pi)^2} K(\mq,\mkk) - \sum_{q_L}\frac{g^2 \gqq
  d^{abc}}{(4\pi)^2} K(\mq,\mkk)\, , \label{eq:gggkkgeneralformexplicitlevicivita}
\end{align}
the expressions for the integrals $I$ and $K$ are given in
equations~\eqref{eq:gggkkfpnastyintegral} and~\eqref{eq:gggkkfpnotquiteasnastyintegral}, respectively, and we emphasise that
the left- and right-handed quark states must be
treated as separate particles in the sum for $A$.

At this stage, we note that the amplitude as calculated so far
contains an anomaly in the current associated with the outgoing
KK~gluon (that is, the on-shell Ward identity
$r^{\rho}F_{\mu\nu\rho}=0$ is not satisfied). This is because we have
taken the gauge $A_4=0$, which we do not have the freedom to do in a
five-dimensional non-Abelian theory with chiral delocalised quarks if
we desire anomaly cancellation~\cite{AnBDK2006}. We must therefore now apply a gauge transformation
that leaves the four-dimensional theory anomaly-free. We note that
from the perspective of our current calculation, this is a technical
requirement that does not affect the final result of the on-shell
calculation. However, it does have the potential to affect the result
for $F_{\mu\nu\rho}$.

\subsection{Cancellation of the anomaly}

It is possible to add to the five-dimensional Lagrangian for this
theory the Chern-Simons term~\cite{Hi2006}
\begin{equation}
\mathcal{L}_{\text{CS}} = c \epsilon^{VWXYZ} \text{Tr}\left( A_V
\partial_W A_X \partial_Y A_Z - \frac{3i}{2} A_V A_W A_X \partial_Y
A_Z - \frac{3}{5} A_V A_W A_X A_Y A_Z \right) ,
\end{equation}
where $A_V = A_V^a t^a$, etc and $V,W,X,\ldots$ are 5-dimensional space-time
indices. 
 Of interest to us are the three-point
$gg\gkk$ interaction vertices that result from this Lagrangian term (the other
vertices will only feature in higher-order corrections). Without
making a specific gauge choice, but keeping the form of the
extra-dimensional wavefunctions noted in Section~6.3 (which we shall
justify momentarily), we obtain from the four-dimensional
perspective two types of interaction term, depending upon whether the
index for the extra dimension attaches to a gauge field or to a
derivative. The terms are the four-dimensional Chern-Simons-like term
\begin{equation}
\mathcal{L}_{\text{CS4}} = c_1
\epsilon^{\mu\nu\rho\gamma}d^{abc}A_{\mu}^{(0)a}
\partial_{\gamma}A_{\nu}^{(0)b} \partial_4 A_{\rho}^{(1)c}
\end{equation}
(we note that the $\partial_4$ can only act on the KK~mode as the
extra-dimensional wavefunctions of the zero modes are flat, and all
other possible terms can be obtained by using symmetry or antisymmetry
arguments from the one given), and the three-point ``axion'' term
\begin{equation}
\mathcal{L}_{\text{Ax}} = c_2
\epsilon^{\mu\nu\gamma\delta}d^{abc}\left(
\partial_{\gamma}A_{\mu}^{a} \partial_{\delta}A_{\nu}^{b} A_4^c +
2 A_{\mu}^{a} \partial_{\gamma}A_{\nu}^{b} \partial_{\delta} A_4^c
\right) .
\end{equation}
The constants $c_1$ and $c_2$ may be chosen separately, since we inherit
two degrees of freedom from the five-dimensional theory, one from the
coefficient of the five-dimensional Chern-Simons term, and one from
the gauge choice. We choose $c_1=0$, and shall choose $c_2$ so as to
provide the requisite anomaly cancellation.

The $A_4$ field also enters into the kinetic gauge term of the
Lagrangian, allowing oscillation between the scalar axion and the
four-dimensional gauge modes.

\begin{figure}[!tbp]
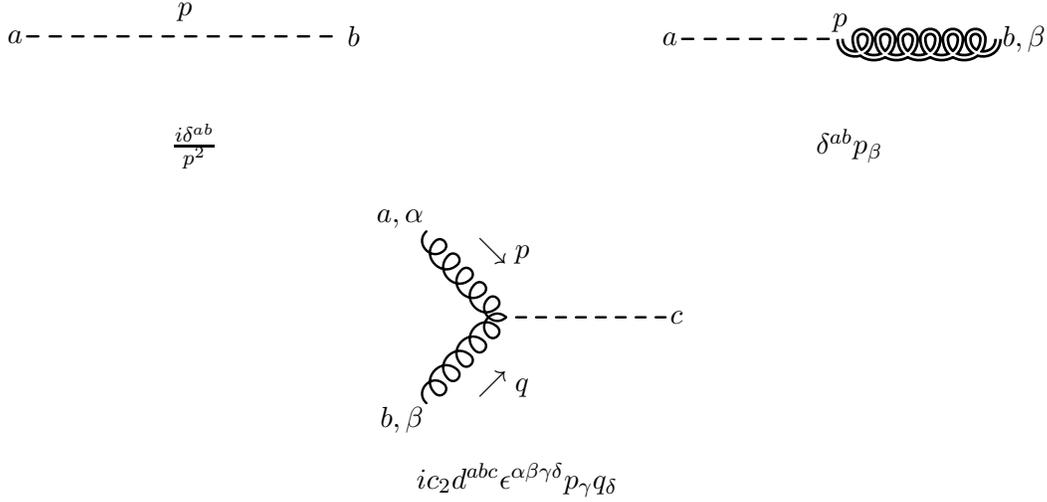

\begin{center}
\feynmanrulesdiagcaptionshort{%
\feynmanrulespropagator{dashes}{a}{b}{p}
}{\frac{i\delta^{ab}}{p^2}}
\hspace{\stretch{1}}
\feynmanrulesdiagcaptionshort{%
\fmfleft{l}
\fmfright{r}
\fmf{dashes,label=$  $,l.side=left}{l,i1}
\fmf{dbl_curly,label=$  $,l.side=left}{i1,r}
\fmfv{label=$ p $,l.dist=1mm,l.a=90}{i1}
\fmfv{label=$ a $,l.dist=0.6mm}{l}
\fmfv{label=$ b,,\beta $,l.dist=0.6mm}{r}
}{\delta^{ab}p_{\beta}}
\\
\feynmanrulesdiagcaption{%
\feynmanrulesthreepoint{dashes}{c}{}{curly}{a,,\alpha}{\searrow
  p}{curly}{b,,\beta}{\nearrow q}
}{ic_2d^{abc}\epsilon^{\alpha\beta\gamma\delta}p_{\gamma}q_{\delta}}
\caption{Additional relevant non-zero Feynman rules for gauges with $A_4 \neq 0$.}\label{fig:chernsimonsadditionalfeynmanrules}
\end{center}
\end{figure}

The additional Feynman rules resulting from the change of gauge and
the addition of the extra Lagrangian term are shown in
Figure~\ref{fig:chernsimonsadditionalfeynmanrules}. %
%
%

\begin{figure}[!tbp]
\begin{center}
\vspace{1.5em}
\begin{fmfgraph*}(\fermiondiagwidth,\fermiondiagheight)
\fmfleft{l1,l2}
\fmfright{r1}
\fmf{gluon}{l2,i1,l1}
\fmf{dashes}{i1,i2}
\fmf{dbl_curly}{i2,r1}
\fmflabel{$\mu,a,p$}{l2}
\fmflabel{$\nu,b,q$}{l1}
\fmflabel{$\rho,c,r$}{r1}
\end{fmfgraph*}\\
\vspace{\fermionspace}
\caption{Additional Feynman diagram for the process $gg\to\gkk$ for
  gauges with $A_4 \neq 0$.}\label{fig:chernsimonsadditionalfeynmandiagram}
\end{center}
\end{figure}
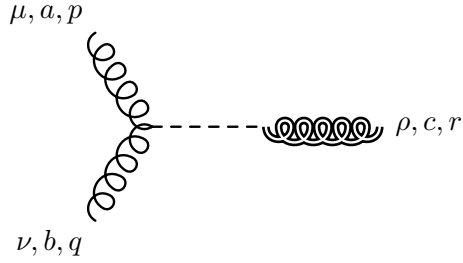

These rules give rise to one extra diagram, shown in
Figure~\ref{fig:chernsimonsadditionalfeynmandiagram}.
%
%

The extra diagram gives a contribution to the
amplitude of
\begin{equation}
F_{\mu\nu\rho}^{(\text{axion})}=
-c_2d^{abc}\frac{1}{r^2}\epsilon_{\mu\nu\gamma\delta}p^{\gamma}q^{\delta}r_{\rho}
\, ,
\end{equation}
and we can choose the constant $c_2$ so that this contribution is equal
on-shell to
\begin{equation}
C\epsilon_{\mu\nu\gamma\delta}p^{\gamma}q^{\delta}r_{\rho}
\, , \label{eq:chernsimonsresultantaddition}
\end{equation}
where $C$ is as given in equation~\eqref{eq:gggkkgeneralformexplicitlevicivita}. This addition means that the
amplitude now satisfies the Ward identity $r^{\rho}F_{\mu\nu\rho}=0$,
without disturbing the other Ward
identities\footnote{To prove the
Ward identity fully it is also necessary to note that the term
\mbox{$A(\eta_{\mu\nu}p\cdot q - q_{\mu}p_{\nu})p_{\rho}$} was derived by
applying the KK~polarisation tensor to the term \mbox{$(A/2)(\eta_{\mu\nu}p\cdot
q - q_{\mu}p_{\nu})(p_{\rho}-q_{\rho})$}.}. Since the term $r_{\rho}$ in
the additional contribution is projected out by the KK~polarisation
tensor, it is clear that this contribution makes no difference to the
square of the final on-shell matrix element. (We recall that the
on-shell $\gkk$ amplitude is given by contracting 
$F_{\mu\nu\rho}$ with the external polarisation vectors.)

Before finishing this technical aside, we note that making a gauge
choice other than $A_4 = 0$ alters the form of the five-dimensional
Yang-Mills equation~\cite{DaHR1999a,Po1999}, and in
particular this has the potential to break the derivation of the
extra-dimensional wavefunction $\chi^{(0)}$ as constant. However, this
alteration may be viewed as a loop-level correction to the equation,
so that the effects of this alteration upon all the diagrams we have
considered so far are two-loop level. This alteration does have the
potential to produce a tree diagram for $gg\to \gkk$ fusion, but the
structure of such a diagram is such that it may be absorbed into the
three-point counterterm diagram and neglected from the perspective of
a separate contribution.

\section{Calculation of the production cross-section \label{sec:sec}}

Our first step in obtaining the production cross-section is to obtain
the square of the matrix element. For this we may use the polarisation
sum formulae
\begin{align}
\gpolp \varepsilon^{\mu'*}(p) &= -\metu{\mu\mu'} + Q^{\mu \mu'}\, ,\\
\gpolq \varepsilon^{\nu'*}(q) &= -\metu{\nu\nu'} + Q^{\nu \nu'}\, ,\\
\gpolrprime \gpolrout &= \frac{r^{\rho}r^{\rho'}}{\mkk^2} -
\metu{\rho\rho'} \, .
\end{align}
where $Q^{\mu \mu'} = (p^{\mu}q^{\mu'}+p^{\mu'}q^{\mu})/(p \cdot q)$
and $Q^{\nu \nu'} = (q^{\nu}p^{\nu'}+q^{\nu'}p^{\nu})/(p \cdot q)$.
The square of the matrix element satisfies
\begin{align}
|\mathcal{M}|^2 &= \gpolp\gpolq
 \gpolrout \gamp
 \varepsilon^{\mu'*}(p)\varepsilon^{\nu'*}(q)\varepsilon_{\gkk}^{\rho'}(r)
 F^*_{\mu'\nu'\rho'} \, ,
\end{align}
and substituting for the polarisation sum formulae and using
equation~\eqref{eq:gggkkgeneralformsimplified}, we
obtain~\cite{special:FORM}
\begin{equation}
|\mathcal{M}|^2 = \frac{\mkk^6}{32}|A|^2 \, ,
\end{equation}
with $A$ as given in equation~\eqref{eq:gggkkgeneralformexplicit}, and
where we have averaged over incoming polarisation states. We
note that the coefficient $C$ has cancelled and we therefore do not
need to evaluate the integrals $K$ that feature in the equation for
coefficient $C$ and not in the equation for coefficient $A$.

We may evaluate $|A|^2$, incorporating averaging over incoming colour
states and summing over outgoing colour states to obtain
\begin{align}
|\mathcal{M}|^2 &= \frac{\mkk^6g^4}{(4\pi)^4}\Bigg|
\frac{4046}{16384}
\gkkksq \left[I(\mkk,\mkk)\right]^2 - \nonumber\\
& \phantom{= \frac{\mkk^6g^2}{32(4\pi)^2}\Bigg\{}
- \frac{51}{256} \gkkk \sum_{q_L,q_R}
   \gqq  [I(\mkk,\mkk)I(\mq,\mkk)] + \nonumber\\
& \phantom{= \frac{\mkk^6g^2}{32(4\pi)^2}\Bigg\{} +\frac{3}{64} \left[
     \sum_{q_L,q_R}\gqq I(\mq,\mkk) \right]^2 \Bigg| 
\, ,
\end{align}
where we have combined the expressions resulting from the real and imaginary parts of the matrix
element to obtain this last expression.

Writing 
\begin{equation}
\vert {\cal M} \vert^2 = {\mkk^6 g^4 \over (4\pi)^4}
\vert \tilde {\cal M} \vert^2 \, ,
\end{equation}
We can write an expression for the cross-section for the production
of on-shell KK gluons from the gluon-initiated states as
\begin{equation}
\sigma = {\mkk^2 \alpha_s^2 \over 8\pi} \int dy x_1 g_a(x_1, \mkk^2) 
x_2 g_b(x_2, \mkk^2) \vert \tilde {\cal M} \vert^2 ,
\end{equation}
where $x_{1,2}=\sqrt{\tau} e^{\pm y}$, with $\sqrt{\tau}=\mkk/\sqrt{s}$,
$y$ being the rapidity of the KK gluon and $\sqrt{s}$ being the total centre
of mass energy of the $pp$ system.

\begin{figure}[htb]
\begin{picture}(4,6)
\put(0,0){\epsfig{file=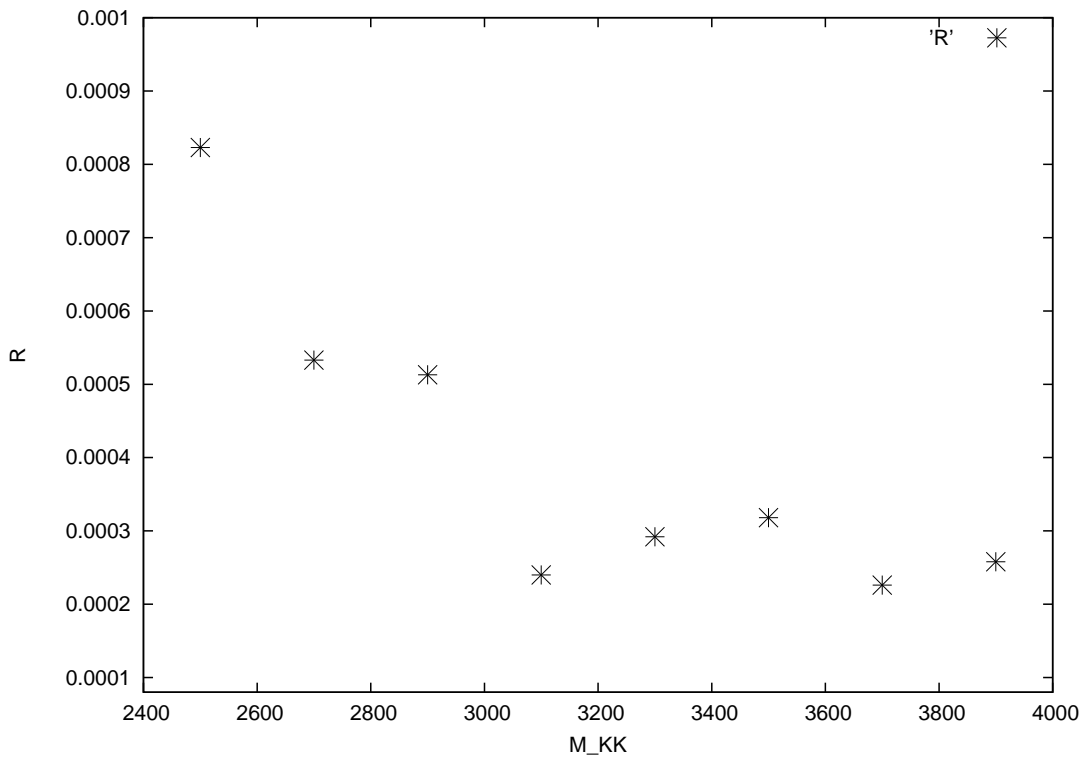, width=4in, angle=-90}}
\end{picture}
\vspace{10cm}
\caption{The ratio of the gluon-initiated NLO and the $q \bar q$ -initiated
LO cross-sections for the production of a KK gluon 
at the LHC with 14TeV centre of mass energy.}
\protect\label{fig7}
\end{figure}

We have used this expression to calculate the cross-section for the KK gluon
from the $gg$-initial state and compared it with the leading order $q \bar q$
result (using the LO cross-section presented in Ref. \cite{GuMS2007}) at
the Large Hadron Collider (LHC), assuming a centre-of-mass energy of 14 TeV.
The ratio is plotted for some typical values of the KK gluon mass
in Fig.~\ref{fig7}. The cross-section from the $gg$ NLO subprocesses turns out to
be less than a thousandth of the LO cross section. This is, in turn, due
to appearance of the large KK gluon mass squared in the denominators of
the integral $I$ which have been analytically studied in Appendix B to 
provide some intuitive basis for these numerical results. 

In principle, to complete the full calculation of the KK gluon cross-section
at NLO one needs to calculate the $q \bar q$-initiated diagrams at NLO\@. But
given that the $gg$-initiated contribution is tiny, it is expected that the
$q \bar q$-initiated contribution will be even smaller due to the suppressed
couplings of valence quarks and the calculation
is, therefore, not of much interest.

\section{Discussion \label{sec:disc}}

As expected, the most significant contribution to the $gg\to\gkk$
production process comes from the $t_R$ loop, which has the strongest
coupling to the $\gkk$ and interferes constructively with other loop
particles. 
Other loops not involving $t_R$ contribute non-negligibly to the process
however, owing partly to the number of different additional loops. 
We note that our final result disagrees with a preliminary result obtained as
part of Ref.~\cite{DjMS2007}. Like Ref.~\cite{DjMS2007}, we also find
that the gg-initiated contribution is negligibly small.

It has been suggested~\cite{DjMS2007} that the decay width for $\gkk
\to t\bar{t}/b\bar{b}$ is sufficiently large to
suggest that  the
narrow width approximation would be an inaccurate approximation to the
total matrix element. In this case, off-shell $\gkk$ effects would be
non-negligible and one would like to generalise our calculation to the
off-shell case. In Appendix~\ref{sec:gen}, we have provided a recipe
to modify the amplitudes presented here to the case where the $\gkk$
is off-shell. However, we should remember that this alone is not enough
for a full study of the off-shell effects for such a study would have to
include the interference effects with the Standard Model. The full
calculation of the interference effects is, however, not of much
interest given the diminutiveness of the effects in the on-shell
case. 

\acknowledgments

This work has been partially supported by STFC\@. We thank G.~Moreau for
communication about the calculation. BCA and JPS would like to thank Howard Haber, Bryan Webber and the members of the Cambridge Supersymmetry Working Group for useful discussions. Part of this work was carried out at the
`Les Houches 2009: Physics at TeV Colliders' and `From Strings to LHC II'
workshops. KS wishes to thank IPPP, Durham and LAPTH, Annecy for 
visits during which this work was completed.

\appendix

\section{Tensors satisfying the general form of Section~\protect\ref{sec:generalform}}\label{sec:tensors_satisfying_general_form}

The following combinations of $p$, $q$, the metric tensor
$\eta$, and the Levi-Civita tensor $\epsilon$, satisfy the conditions
given in Section~\ref{sec:generalform} that must be satisfied by the tensor
$t^1_{\mu\nu\rho\alpha\beta}$, as defined by
equation~\eqref{eq:gggeneralformtensor}. Degenerate combinations have
been removed (so, for example, terms with a $q_{\rho}$ are replaced by
terms with $-p_{\rho}$, and terms with more than one $\epsilon$ tensor are
removed in favour of metric tensors). Each term may have a scalar
coefficient.
\begin{gather}
\met{\alpha\beta}p_{\nu}q_{\mu}p_{\rho} -
  \met{\mu\beta}p_{\nu}q_{\alpha}p_{\rho} -
  \met{\alpha\nu}p_{\beta}q_{\mu}p_{\rho} +
  \met{\mu\nu}p_{\beta}q_{\alpha}p_{\rho}\\
\met{\alpha\beta}\met{\mu\nu}p_{\rho} -
  \met{\mu\beta}\met{\alpha\nu}p_{\rho}\\
\met{\alpha\beta}\met{\mu\rho}p_{\nu} -
  \met{\mu\beta}\met{\alpha\rho}p_{\nu} -
  \met{\alpha\nu}\met{\mu\rho}p_{\beta} +
  \met{\mu\nu}\met{\alpha\rho}p_{\beta}\\
\met{\beta\alpha}\met{\nu\rho}q_{\mu} -
  \met{\nu\alpha}\met{\beta\rho}q_{\mu} -
  \met{\beta\mu}\met{\nu\rho}q_{\alpha} +
  \met{\nu\mu}\met{\beta\rho}q_{\alpha}\displaybreak[0]\\
\eps{\mu\nu\alpha\beta}p_{\rho}\displaybreak[0]\\
\eps{\mu\nu\rho\alpha}p_{\beta} -
  \eps{\mu\beta\rho\alpha}p_{\nu}\displaybreak[0]\\
\eps{\mu\nu\rho\beta}q_{\alpha} -
  \eps{\alpha\nu\rho\beta}q_{\mu}\displaybreak[0]\\
\eps{\mu\nu\rho\gamma}p^{\gamma}p_{\beta}q_{\alpha} -
  \eps{\alpha\nu\rho\gamma}p^{\gamma}p_{\beta}q_{\mu} -
  \eps{\mu\beta\rho\gamma}p^{\gamma}p_{\nu}q_{\alpha} +
  \eps{\alpha\beta\rho\gamma}p^{\gamma}p_{\nu}q_{\mu}\displaybreak[0]\\
\eps{\mu\nu\rho\gamma}q^{\gamma}p_{\beta}q_{\alpha} -
  \eps{\alpha\nu\rho\gamma}q^{\gamma}p_{\beta}q_{\mu} -
  \eps{\mu\beta\rho\gamma}q^{\gamma}p_{\nu}q_{\alpha} +
  \eps{\alpha\beta\rho\gamma}q^{\gamma}p_{\nu}q_{\mu}\displaybreak[0]\\
\eps{\mu\nu\rho\gamma}p^{\gamma}\met{\alpha\beta} -
  \eps{\alpha\nu\rho\gamma}p^{\gamma}\met{\mu\beta} -
  \eps{\mu\beta\rho\gamma}p^{\gamma}\met{\nu\alpha} +
  \eps{\alpha\beta\rho\gamma}p^{\gamma}\met{\nu\mu}\displaybreak[0]\\
\eps{\mu\nu\rho\gamma}q^{\gamma}\met{\alpha\beta} -
  \eps{\alpha\nu\rho\gamma}q^{\gamma}\met{\mu\beta} -
  \eps{\mu\beta\rho\gamma}q^{\gamma}\met{\nu\alpha} +
  \eps{\alpha\beta\rho\gamma}q^{\gamma}\met{\nu\mu}\displaybreak[0]\\
\eps{\mu\nu\alpha\gamma}p^{\gamma}p_{\rho}p_{\beta} -
  \eps{\mu\beta\alpha\gamma}p^{\gamma}p_{\rho}p_{\nu}\displaybreak[0]\\
\eps{\mu\nu\alpha\gamma}q^{\gamma}p_{\rho}p_{\beta} -
  \eps{\mu\beta\alpha\gamma}q^{\gamma}p_{\rho}p_{\nu}\displaybreak[0]\\
\eps{\mu\nu\alpha\gamma}p^{\gamma}\met{\rho\beta} -
  \eps{\mu\beta\alpha\gamma}p^{\gamma}\met{\rho\nu}\displaybreak[0]\\
\eps{\mu\nu\alpha\gamma}q^{\gamma}\met{\rho\beta} -
  \eps{\mu\beta\alpha\gamma}q^{\gamma}\met{\rho\nu}\displaybreak[0]\\
\eps{\mu\nu\beta\gamma}p^{\gamma}p_{\rho}q_{\alpha} -
  \eps{\alpha\nu\beta\gamma}p^{\gamma}p_{\rho}q_{\mu}\displaybreak[0]\\
\eps{\mu\nu\beta\gamma}q^{\gamma}p_{\rho}q_{\alpha} -
  \eps{\alpha\nu\beta\gamma}q^{\gamma}p_{\rho}q_{\mu}\displaybreak[0]\\
\eps{\mu\nu\beta\gamma}p^{\gamma}\met{\rho\alpha} -
  \eps{\alpha\nu\beta\gamma}p^{\gamma}\met{\rho\mu}\displaybreak[0]\\
\eps{\mu\nu\beta\gamma}q^{\gamma}\met{\rho\alpha} -
  \eps{\alpha\nu\beta\gamma}q^{\gamma}\met{\rho\mu}\displaybreak[0]\\
\eps{\mu\nu\gamma\delta}p^{\gamma}q^{\delta}p_{\rho}q_{\alpha}p_{\beta}
  -
  \eps{\alpha\nu\gamma\delta}p^{\gamma}q^{\delta}p_{\rho}q_{\mu}p_{\beta}
  -
  \eps{\mu\beta\gamma\delta}p^{\gamma}q^{\delta}p_{\rho}q_{\alpha}p_{\nu}
  +
  \eps{\alpha\beta\gamma\delta}p^{\gamma}q^{\delta}p_{\rho}q_{\mu}p_{\nu}\displaybreak[0]\\
\eps{\mu\nu\gamma\delta}p^{\gamma}q^{\delta}p_{\rho}\met{\alpha\beta}
  -
  \eps{\alpha\nu\gamma\delta}p^{\gamma}q^{\delta}p_{\rho}\met{\mu\beta}
  -
  \eps{\mu\beta\gamma\delta}p^{\gamma}q^{\delta}p_{\rho}\met{\alpha\nu}
  +
  \eps{\alpha\beta\gamma\delta}p^{\gamma}q^{\delta}p_{\rho}\met{\mu\nu}\displaybreak[0]\\
\eps{\mu\nu\gamma\delta}p^{\gamma}q^{\delta}\met{\rho\alpha}p_{\beta}
  -
  \eps{\alpha\nu\gamma\delta}p^{\gamma}q^{\delta}\met{\rho\mu}p_{\beta} -
  \eps{\mu\beta\gamma\delta}p^{\gamma}q^{\delta}\met{\rho\alpha}p_{\nu} +
  \eps{\alpha\beta\gamma\delta}p^{\gamma}q^{\delta}\met{\rho\mu}p_{\nu}\displaybreak[0]\\
\eps{\mu\nu\gamma\delta}p^{\gamma}q^{\delta}\met{\rho\beta}q_{\alpha}
  -
  \eps{\alpha\nu\gamma\delta}p^{\gamma}q^{\delta}\met{\rho\beta}q_{\mu} -
  \eps{\mu\beta\gamma\delta}p^{\gamma}q^{\delta}\met{\rho\nu}q_{\alpha} +
  \eps{\alpha\beta\gamma\delta}p^{\gamma}q^{\delta}\met{\rho\nu}q_{\mu}\displaybreak[0]\\
\eps{\rho\beta\gamma\delta}p^{\gamma}q^{\delta}\met{\mu\nu}q_{\alpha}
  -
  \eps{\rho\beta\gamma\delta}p^{\gamma}q^{\delta}\met{\alpha\nu}q_{\mu} -
  \eps{\rho\nu\gamma\delta}p^{\gamma}q^{\delta}\met{\mu\beta}q_{\alpha} +
  \eps{\rho\nu\gamma\delta}p^{\gamma}q^{\delta}\met{\alpha\beta}q_{\mu}\\
\eps{\rho\alpha\gamma\delta}p^{\gamma}q^{\delta}\met{\beta\mu}p_{\nu}
  -
  \eps{\rho\mu\gamma\delta}p^{\gamma}q^{\delta}\met{\beta\alpha}p_{\nu} - 
  \eps{\rho\alpha\gamma\delta}p^{\gamma}q^{\delta}\met{\nu\mu}p_{\beta} +
  \eps{\rho\mu\gamma\delta}p^{\gamma}q^{\delta}\met{\nu\alpha}p_{\beta}\\
\eps{\alpha\mu\gamma\delta}p^{\gamma}q^{\delta}\met{\rho\beta}p_{\nu}
  -
  \eps{\alpha\mu\gamma\delta}p^{\gamma}q^{\delta}\met{\rho\nu}p_{\beta}\\
\eps{\beta\nu\gamma\delta}p^{\gamma}q^{\delta}\met{\rho\alpha}q_{\mu}
  -
  \eps{\beta\nu\gamma\delta}p^{\gamma}q^{\delta}\met{\rho\mu}q_{\alpha}
\end{gather}

\section{Analytic evaluation of the Feynman parameter integral used in the calculation\label{sec:gggkkfpintegralanalytic}}

In the calculation we arrive at the integral (equation~\eqref{eq:gggkkfpnastyintegral})
\begin{equation}
I(\mq, \mkk) = \dblfeynint \frac{xy(1-x-y)}{\mq^2 -
  xy\mkk^2} \, .
\end{equation}
We note that in the integration region the expression $xy$ has a
maximum value of $1/4$, so for Standard Model quarks (where $\mkk \gg
2\mq$) we may approximate the integral by $I(0,\mkk)$, which gives us
\begin{equation}
I(\mq, \mkk) \approx - \frac{1}{6\mkk^2} \, .
\end{equation}

We also obtain the integral $I(\mkk,\mkk)$.
This may be evaluated as follows:
\begin{align}
 \dblfeynint & \frac{xy(1-x-y)}{\mq^2-xy\mkk^2} = \nonumber\\
= \frac{1}{\mkk^2} \dblfeynint & (x+y-1) + \frac{1-x-y}{1-xy} =
\nonumber\\
= \frac{1}{\mkk^2} \dblfeynint & (2x-1) + \frac{1-2x}{1-xy} \, ,
\end{align}
{since the integration region is symmetrical about the line
  $x=y$ so we may interchange $x$ and $y$ in any term in the
  integrand. We may evaluate this to obtain}
\begin{align}
& -\frac{1}{6\mkk^2} + \frac{1}{\mkk^2} \sglfeynint
\left[\frac{(2x-1)}{x}\log (1-xy) \right]_{y=0}^{y=1-x} = \nonumber\\
= & -\frac{1}{6\mkk^2} + \frac{1}{\mkk^2} \sglfeynint
\left(2-\frac{1}{x}\right) \log(x^2-x+1) = \nonumber\\
= & -\frac{1}{6\mkk^2} + \frac{1}{\mkk^2} \sglfeynint
\left(2-\frac{1}{x}\right) \left[
  \log\left(x-\frac{1}{2}-i\frac{\sqrt{3}}{2}\right) +
  \log\left(x-\frac{1}{2}+i\frac{\sqrt{3}}{2}\right) \right] = \nonumber\\
= & -\frac{1}{6\mkk^2} +
\frac{2}{\mkk^2}\left(-2+\frac{\pi}{\sqrt{3}}\right)-
\nonumber\\ & -\frac{1}{\mkk^2} \sglfeynint
\frac{1}{x} \left[
  \log\left(1-\frac{x}{\frac{1}{2}+i\frac{\sqrt{3}}{2}}\right) +
  \log\left(1-\frac{x}{\frac{1}{2}+i\frac{\sqrt{3}}{2}}\right) \right] =
\nonumber\\
= & \frac{1}{\mkk^2}\left(\frac{2\pi}{\sqrt{3}}-\frac{25}{6} \right) +
\frac{1}{\mkk^2}\left[
  \li_2\left(\frac{1}{\frac{1}{2}+i\frac{\sqrt{3}}{2}}\right)+
  \li_2\left(\frac{1}{\frac{1}{2}-i\frac{\sqrt{3}}{2}}\right) \right] =
\nonumber\\
= & \frac{1}{\mkk^2}\left(\frac{2\pi}{\sqrt{3}}-\frac{25}{6} \right) +
\frac{1}{\mkk^2}\left[
  \li_2\left({\frac{1}{2}-i\frac{\sqrt{3}}{2}}\right)+
  \li_2\left({\frac{1}{2}+i\frac{\sqrt{3}}{2}}\right) \right] =
\nonumber\\
= & \frac{1}{\mkk^2}\left(\frac{2\pi}{\sqrt{3}}-\frac{25}{6} \right) +
\frac{1}{\mkk^2}\left[ \frac{\pi^2}{6} -
  \log\left(\frac{1}{2}-i\frac{\sqrt{3}}{2}\right)\log\left(\frac{1}{2}+i\frac{\sqrt{3}}{2}\right)
  \right] =
\nonumber\\
= & \frac{1}{\mkk^2}\left(\frac{2\pi}{\sqrt{3}}-\frac{25}{6} \right) +
\frac{1}{\mkk^2}\left[ \frac{\pi^2}{6} -
 \frac{\pi^2}{9}
  \right] = \nonumber\\
= &
\frac{1}{\mkk^2}\left(\frac{\pi^2}{18}+\frac{2\pi}{\sqrt{3}}-\frac{25}{6}
\right) \approx \frac{1}{100\mkk^2} ,
\end{align}
where we have used a number of standard properties of the dilogarithm
function (see, e.g.~Appendix~E.2 of reference~\cite{book:field}).

A similar approach may be used to obtain a full analytic solution to
the integral $I(\mq,\mkk)$, although because the dilogarithms do not
simplify as usefully in that case the utility of the overall solution
is comparatively smaller.

\section{Generalisation of our results to off-shell KK gluon production\label{sec:gen}}
 We note that the argument of
 Section~\ref{sec:generalform} only assumes that the KK~gluon is
 on-shell in allowing the replacement $p_{\rho} \leftrightarrow
 -q_{\rho}$, so the argument still holds with the exception that we
 obtain additional possible terms in the general form of the amplitude,
 which correspond to replacing $p_{\rho}$ with $q_{\rho}$ in
 equations~(\ref{eq:gggkkgenformone})
 and~(\ref{eq:gggkkgenformtwo}). The symmetry of the diagrams means we
 know that the terms we shall obtain will replace $p_{\rho}$ with
 $(p_{\rho}-q_{\rho})/2$, although when calculating the diagrams we
 also have to avoid making the replacements $2p\cdot q = r^2 =
 \mkk^2$. In addition, because we cannot use the polarisation tensor $\gpolr$
 when deriving the general form of the amplitude, we would have to include the
 term derived in equation~\eqref{eq:chernsimonsresultantaddition}. 

\bibliographystyle{JHEP}
\bibliography{abbrev_short,general_20090622}

\end{fmffile}
\end{document}